\documentclass[iop, apj, twocolappendix, numberedappendix]{emulateapj}

\usepackage{xspace}
\usepackage{amsmath}
\usepackage{framed} 
\usepackage{txfonts}
\usepackage{epstopdf} 
\usepackage{color}
\usepackage{rotating}
\usepackage{natbib}
\usepackage{ulem}
\usepackage{xspace}

\usepackage[colorlinks=true,urlcolor=blue,linkcolor=blue,citecolor=blue]{hyperref}

\journalinfo{The Astrophysical Journal, {\rm 799:108 (16pp), 2015 January 20}}
\submitted{Received 2014 August 6; accepted 2014 November 17; published 2015 January 20}

\newcommand{\kpch}{\>{h^{-1}{\rm kpc}}}
\newcommand{\mpch}{\>h^{-1}{\rm {Mpc}}}

\newcommand{\msunh}{\>h^{-1}\rm M_\odot}

\newcommand{\cm}{\>{\rm cm}}

\def\LCDM{$\Lambda$CDM\ }

\def\mvir{M_{\rm vir}}
\def\rvir{R_{\rm vir}}
\def\gcm3{\mathrm{g} / \mathrm{cm}^3}

\def\gtsima{$\; \buildrel > \over \sim \;$}
\def\ltsima{$\; \buildrel < \over \sim \;$}
\def\prosima{$\; \buildrel \propto \over \sim \;$}
\def\gsim{\lower.7ex\hbox{\gtsima}}
\def\lsim{\lower.7ex\hbox{\ltsima}}
\def\simgt{\lower.7ex\hbox{\gtsima}}
\def\simlt{\lower.7ex\hbox{\ltsima}}
\def\simpr{\lower.7ex\hbox{\prosima}}

\newcommand{\avg}[1]{\langle #1 \rangle}

\def\cm{$c$-$M$\xspace}
\def\cnu{$c$-$\nu$\xspace}
\def\cmr{$c$-$M$ relation\xspace}
\def\cnur{$c$-$\nu$ relation\xspace}
\def\cmrs{$c$-$M$ relations\xspace}
\def\cnurs{$c$-$\nu$ relations\xspace}
\def\ctocnur{$\ctoc$-$\nu$ relation\xspace}

\def\xcurv{$\avg{x}$\xspace}

\def\rs{r_{\rm s}}

\def\rhoc{\rho_{\rm c}}
\def\rhom{\rho_{\rm m}}

\def\cvir{c_{\rm vir}}
\def\nuvir{\nu_{\rm vir}}

\def\mtom{M_{\rm 200m}}
\def\rtom{R_{\rm 200m}}
\def\ctom{c_{\rm 200m}}

\def\mtoc{M_{\rm 200c}}
\def\rtoc{R_{\rm 200c}}
\def\ctoc{c_{\rm 200c}}
\def\nutoc{\nu_{\rm 200c}}

\def\rfoc{R_{\rm 500c}}
\def\cfoc{c_{\rm 500c}}

\usepackage{etoolbox}
\makeatletter
\patchcmd{\NAT@citex}
  {\@citea\NAT@hyper@{\NAT@nmfmt{\NAT@nm}\NAT@date}}
  {\@citea\NAT@nmfmt{\NAT@nm}\NAT@hyper@{\NAT@date}}
  {}% Do nothing if patch works
  {}% Do nothing if patch fails
\patchcmd{\NAT@citex}
  {\@citea\NAT@hyper@{%
     \NAT@nmfmt{\NAT@nm}%
     \hyper@natlinkbreak{\NAT@aysep\NAT@spacechar}{\@citeb\@extra@b@citeb}%
     \NAT@date}}
  {\@citea\NAT@nmfmt{\NAT@nm}%
   \NAT@aysep\NAT@spacechar%
   \NAT@hyper@{\NAT@date}}
  {}% Do nothing if patch works
  {}% Do nothing if patch fails
\patchcmd{\NAT@citex}
  {\@citea\NAT@hyper@{%
     \NAT@nmfmt{\NAT@nm}%
     \hyper@natlinkbreak{\NAT@spacechar\NAT@@open\if*#1*\else#1\NAT@spacechar\fi}%
       {\@citeb\@extra@b@citeb}%
     \NAT@date}}
  {\@citea\NAT@nmfmt{\NAT@nm}%
   \NAT@spacechar\NAT@@open\if*#1*\else#1\NAT@spacechar\fi%
   \NAT@hyper@{\NAT@date}}
  {}% Do nothing if patch works
  {}% Do nothing if patch fails
\makeatother

\shorttitle{a universal model for halo concentrations}
\shortauthors{Diemer \& Kravtsov}
\begin{document}

%%%%%%%%%%%%%%%%%%%%%%%%%%%%%%%%%%%%%%%%%%%%%%%%%%%%%%%%%%%%%%%%%%%%%%%%%%
% TITLE ETC
%%%%%%%%%%%%%%%%%%%%%%%%%%%%%%%%%%%%%%%%%%%%%%%%%%%%%%%%%%%%%%%%%%%%%%%%%%

\title{A universal model for halo concentrations}

\author{Benedikt Diemer \altaffilmark{1,2} and Andrey V. Kravtsov\altaffilmark{1,2,3}}

\affil{
$^1$ Department of Astronomy and Astrophysics, The University of Chicago, Chicago, IL 60637, USA; \href{mailto:bdiemer@oddjob.uchicago.edu}{bdiemer@oddjob.uchicago.edu} \\
$^2$ Kavli Institute for Cosmological Physics, The University of Chicago, Chicago, IL 60637, USA \\
$^3$ Enrico Fermi Institute, The University of Chicago, Chicago, IL 60637, USA \\
}

%%%%%%%%%%%%%%%%%%%%%%%%%%%%%%%%%%%%%%%%%%%%%%%%%%%%%%%%%%%%%%%%%%%%%%%%%%
% ABSTRACT
%%%%%%%%%%%%%%%%%%%%%%%%%%%%%%%%%%%%%%%%%%%%%%%%%%%%%%%%%%%%%%%%%%%%%%%%%%

\begin{abstract}
We present a numerical study of dark matter halo concentrations in $\Lambda$CDM and self-similar cosmologies. We show that the relation between concentration, $c$, and peak height, $\nu$, exhibits the smallest deviations from universality if halo masses are defined with respect to the critical density of the universe. These deviations can be explained by the residual dependence of concentration on the local slope of the matter power spectrum, $n$, which affects both the normalization and shape of the \cnur. In particular, there is no well-defined floor in the concentration values. Instead, the minimum concentration depends on redshift: at fixed $\nu$, halos at higher $z$ experience steeper slopes $n$, and thus have lower minimum concentrations. We show that the concentrations in our simulations can be accurately described by a universal seven-parameter function of only $\nu$ and $n$. This model matches our $\Lambda$CDM results to $\lesssim 5\%$ accuracy up to $z = 6$, and matches scale-free $\Omega_{\rm m} = 1$ models to $\lesssim 15\%$. The model also reproduces the low concentration values of Earth--mass halos at $z \approx 30$, and thus correctly extrapolates over $16$ orders of magnitude in halo mass. The predictions of our model differ significantly from all models previously proposed in the literature at high masses and redshifts. Our model is in excellent agreement with recent lensing measurements of cluster concentrations.
\end{abstract}


%--------------------------------------------------------------------------------------------
\section{Introduction}
\label{sec:intro}
%--------------------------------------------------------------------------------------------

A theoretical understanding of the density structure of halos is essential for the correct interpretation of a variety of astronomical observations, from the kinematics of stars and gas in galaxies to the lensing signal in the outskirts of galaxies and clusters \citep[see, e.g.,][for a recent review]{courteau_14}. Thus, studies of the radial density profiles of halos were one of the first applications of cosmological simulations. Early simulations indicated that the density profiles forming in a cold dark matter (CDM) scenario are close to the isothermal profile, $\rho\propto r^{-2}$ \citep{frenk_88_haloformation}, while subsequent higher-resolution simulations showed that the profiles have a slope that slowly changes from steep values of about $-3$ to $-4$ around the virial radius to shallower slopes of about $-1$ in the inner regions \citep{dubinski_91, navarro_95, cole_96_halostructure}. Nevertheless, the radial range where the density profile slope is close to the isothermal value of $-2$ is highly important, given that such a mass distribution is needed to explain the flat rotation curves of galaxies.

One of the most widely used parameters characterizing the density profiles of halos is thus {\it concentration}, defined as the ratio of the outer ``virial'' radius and the radius at which the logarithmic slope is $-2$. This definition applies to any form of the profiles, including common analytic functions, such as the Hernquist profile \citep{hernquist_90, dubinski_91}, the profile proposed by \citet[][hereafter NFW]{navarro_95, navarro_96, navarro_97_nfw}, or the Einasto profile \citep{einasto_65, einasto_69, navarro_04}. For the NFW profile, in particular, the concentration of a halo with a given virial mass fully specifies its profile.

Given the importance of concentration, its calibration has been the subject of numerous studies. \citet{navarro_96, navarro_97_nfw} first suggested a model in which concentration depends on the epoch at which a certain fraction of a halo's mass has been assembled. Although the original model was shown to predict an incorrect evolution of concentrations \citep{bullock_01_halo_profiles}, the general idea that concentration is related to a halo's assembly history was shown to be valid by subsequent studies. \citet[][see also \citealt{wechsler_02_halo_assembly} and \citealt{zhao_03_concentration}]{bullock_01_halo_profiles} showed that, during the evolutionary stage when the mass accretion rate of a halo slows down, its scale radius remains approximately constant, and thus concentration scales as the virial radius, $c\propto R_{\rm vir}$. In the regime of a high mass accretion rate, on the other hand, the scale radius scales approximately as the virial radius and $c \approx \rm const$ \citep{zhao_03_concentration}. Overall, there was found to be a tight relation between concentration, the shape of the profile at any given time, and the mass assembly history (MAH) of the main halo progenitor prior to that time \citep{zhao_03_concentration, zhao_09_mah, ludlow_14_cm}. The MAH depends on the amplitude and shape of the initial density peak \citep[e.g.,][]{dalal_08}, which, in turn, depend on the mass scale of the peak as well as on the parameters of the background cosmological model. Thus, halo concentrations depend on mass, redshift, and cosmological parameters. 

Many theoretical studies have calibrated these dependencies using cosmological simulations of halos \citep[e.g.,][]{avilareese_99, jing_00_profiles2,bullock_01_halo_profiles, colin_04, dolag_04, neto_07, duffy_08, gao_08, maccio_08, klypin_11_bolshoi, munozcuartas_11, bhattacharya_13, dutton_14}, usually approximating the relation between concentration and mass (or peak height) as a power-law. The main limitation of such parameterized fits to simulation results is that they generally cannot be extrapolated beyond the cosmological model, mass, and redshift range for which they were calibrated. 

For this reason, a number of more general models for halo concentrations have been proposed, many of them based on the tight connection between halo concentration and formation epoch \citep{navarro_97_nfw, bullock_01_halo_profiles, eke_01_concentrations, zhao_09_mah, giocoli_12}. Given the influence of the parameters characterizing the initial density peak on a halo's MAH \citep{dalal_08}, one could expect that concentration should be a function of such peak parameters. Indeed, numerical studies have shown that much of the dependence of concentration on cosmology and redshift can be taken into account by expressing concentrations as a function of peak height, $\nu$ \citep[see Equation (\ref{eq:nu}) below;][]{zhao_09_mah, prada_12, ludlow_14_cm, dutton_14}. At the same time, these studies showed that the \cnur is not quite universal with redshift, prompting \citet{prada_12} to present a fitting formula which parameterizes the extra dependency using an arbitrary time rescaling function. This model, however, fails for non-$\Lambda$CDM cosmologies, as we show in Section \ref{sec:discussion:comp2}.

Given the general logic that halo concentration and MAH should depend on the properties of the initial density peak, it is natural to interpret the non-universality of the \cnur as an indication that there is at least one additional peak variable that controls concentration. In this paper, we quantify deviations of the \cnur from universality for different choices of the ``virial'' radius, and explore the possible second parameter controlling concentration. We identify the local slope of the power spectrum, $n$, as such a parameter, and present an accurate, universal model in which halo concentrations depend only on $\nu$ and $n$. We note that either an explicit or implicit dependence of concentration on the power spectrum slope is also included in the models of \citet{eke_01_concentrations}, \citet{bullock_01_halo_profiles}, and \citet{zhao_09_mah}. However, the specifics of these models, particularly the ways in which the dependence on the power spectrum slope is modeled, differ significantly from our model (Section \ref{sec:discussion}). 

We demonstrate that our model accurately describes concentrations in both $\Lambda$CDM cosmologies with different parameters and self-similar cosmologies with power-law spectra and $\Omega_{\rm m} = 1$. We also show that the model provides reasonably accurate predictions for Earth--mass halos at $z = 30$, far outside the mass and redshift regime in which the model was calibrated. Although ultimately the concentrations of realistic halos are impacted by baryonic effects which are still rather uncertain \citep{rudd_08, duffy_10, velliscig_14}, the results presented in this paper demonstrate that a simple, universal baseline model of concentrations does exist. While we focus on the NFW approximation to halo profiles, our conclusions should be general and applicable to concentrations defined for other analytical profiles. 

The paper is organized as follows. In Section \ref{sec:numerical:definitions} we describe various definitions of halo mass, radius, and peak height, as well as the corresponding notation used in this paper. In Section \ref{sec:numerical} we describe our numerical simulations and halo sample selection criteria, as well as our halo finder and the method it uses to estimate concentrations. In Section \ref{sec:results} we present our results, in particular an accurate model for concentrations, which we compare to previous studies in Section \ref{sec:discussion}. We summarize our conclusions in Section \ref{sec:conclusion}. In the Appendix we discuss the redshift evolution of the concentrations of individual halos (Appendix \ref{sec:discussion:evolution}), peak curvature as a candidate for a second parameter controlling concentration (Appendix \ref{sec:discussion:curvature}), as well as the conversion or our model to other mass definitions (Appendix \ref{sec:discussion:mdefs}). We provide a stand--alone Python module to compute the predictions of our concentration model for arbitrary cosmologies at \url{benediktdiemer.com/code}.

%--------------------------------------------------------------------------------------------
\section{Mass and Peak Height Definitions}
\label{sec:numerical:definitions}
%--------------------------------------------------------------------------------------------

Throughout this paper, we denote the mean matter density of the universe as $\rho_{\rm m}$, and the critical density as $\rho_{\rm c}$. Spherical overdensity mass is defined as the mass within the radius enclosing a given density contrast $\Delta$ relative  to $\rho_{\rm m}$ or $\rho_{\rm c}$ at a particular redshift, such that
\begin{equation}
\label{eq:mdelta_m}
M_{\Delta \rm m} = M(<R_{\Delta \rm m})= \frac{4 \pi}{3} \Delta\rho_{\rm m}(z)R^3_{\Delta {\rm m}} \,,
\end{equation}
e.g. $\mtom$, or 
\begin{equation}
\label{eq:mdelta_c}
M_{\Delta \rm c} = M(<R_{\Delta \rm c})= \frac{4 \pi}{3} \Delta\rho_{\rm c}(z)R^3_{\Delta {\rm c}} \,,
\end{equation}
e.g. $\mtoc$. We reserve the labels $\mvir$ and $\rvir$ for a varying contrast $\Delta_{\rm vir}(z)$ which we compute using the approximation of \citet{bryan_98_virial}. The concentration of a halo is defined as the ratio of the virial radius to the scale radius, $\rs$,
\begin{equation}
c_{\Delta} \equiv R_{\Delta} / \rs \,,
\end{equation}
where $c$ carries the same label as $R$, such as $\ctoc$. In our analysis, we often express halo mass as peak height, $\nu$, which is defined as
\begin{equation}
\nu \equiv \frac{\delta_{\rm c}}{\sigma(M, z)} = \frac{\delta_{\rm c}}{\sigma(M, z = 0) \times D_+(z)},
\label{eq:nu}
\end{equation}
where $\delta_{\rm c} = 1.686$ is the critical overdensity for collapse derived from the spherical top hat collapse model \citep[][we ignore a weak dependence of $\delta_{\rm c}$ on cosmology and redshift]{gunn_72_sphericalcollapse}, and $D_+(z)$ is the linear growth factor normalized to unity at $z=0$.  Here $\sigma$ is the rms density fluctuation in a sphere of radius, 
\begin{equation}
\label{eq:sigma}
\sigma^2(R,z) = \frac{1}{2 \pi^2} \int_0^{\infty} k^2 P(k,z) |\tilde{W}(kR)|^2 dk \,,
\end{equation}
where mass and radius are defined as
\begin{equation}
\label{eq:mtor}
M = (4 \pi/3) \rho_{\rm m}(z=0) R^3 \,.
\end{equation}
Here $\tilde{W}(kR)$ is the Fourier transform of the spherical top hat filter function, and $P(k,z)=D_+^2(z)P(k,0)$ is the linear matter power spectrum. We use the accurate approximation of \citet{eisenstein_98} to compute $P(k)$, normalized such that $\sigma(8 \mpch) = \sigma_8$.

As with the mass and radius definitions above, we use $\nu_{\Delta}$ to denote the peak height defined by setting $M = M_{\Delta}$ in Equation (\ref{eq:mtor}). Finally, the characteristic non--linear mass, $M^*$, is defined as the mass where $\sigma(M^*) = \delta_{\rm c}$, and thus $\nu(M^*) = 1$.

%--------------------------------------------------------------------------------------------
\section{Numerical Simulations and Methods}
\label{sec:numerical}
%--------------------------------------------------------------------------------------------

\begin{deluxetable*}{lccccccccl}
\tablecaption{$N$-body Simulations
\label{table:sims}}
\tablewidth{0pt}
\tablehead{
\colhead{Name} &
\colhead{$L\, (\mpch)$} &
\colhead{$N^3$} &
\colhead{$m_{\rm p}\, (\msunh)$} &
\colhead{$\epsilon\, (\kpch)$} &
\colhead{$\epsilon / (L / N)$} &
\colhead{$z_{\rm initial}$} &
\colhead{$z_{\rm final}$} &
\colhead{Cosmology} &
\colhead{Reference}
}
\startdata
L2000 & $2000$ & $1024^3$ & $5.6 \times 10^{11}$  & $65$  & $1/30$ & $49$ & $0$ & $WMAP$ (Bolshoi) & This paper \\
L1000 & $1000$ & $1024^3$ & $7.0 \times 10^{10}$ & $33$ & $1/30$ & $49$ & $0$ & $WMAP$ (Bolshoi) & \citealt{diemer_13_scalingrel} \\
L0500 & $500$  & $1024^3$ & $8.7 \times 10^{9}$  & $14$ & $1/35$  & $49$ & $0$ & $WMAP$ (Bolshoi) & \citealt{diemer_14_profiles} \\
L0250 & $250$  & $1024^3$ & $1.1 \times 10^{9}$  & $5.8$  & $1/42$  & $49$ & $0$ & $WMAP$ (Bolshoi) & \citealt{diemer_14_profiles} \\
L0125 & $125$  & $1024^3$ & $1.4 \times 10^{8}$  & $2.4$  & $1/51$  & $49$ & $0$ & $WMAP$ (Bolshoi) & \citealt{diemer_14_profiles} \\
L0063 & $62.5$ & $1024^3$ & $1.7 \times 10^{7}$  & $1.0$  & $1/60$ & $49$ & $0$ & $WMAP$ (Bolshoi) & \citealt{diemer_14_profiles} \\
L0031 & $31.25$ & $1024^3$ & $2.1 \times 10^{6}$  & $0.25$  & $1/122$ & $49$ & $2$ & $WMAP$ (Bolshoi) & This paper \\
L0500-Planck & $500$  & $1024^3$ & $1.0 \times 10^{10}$  & $14$ & $1/35$  & $49$ & $0$ & Planck & This paper \\
L0250-Planck & $250$  & $1024^3$ & $1.3 \times 10^{9}$  & $5.8$  & $1/42$  & $49$ & $0$ & Planck & This paper \\
L0125-Planck & $125$  & $1024^3$ & $1.6 \times 10^{8}$  & $2.4$  & $1/51$  & $49$ & $0$ & Planck & This paper \\
L0250-High-$\sigma_8$ & $250$  & $1024^3$ & $1.1 \times 10^{9}$  & $5.8$  & $1/42$  & $49$ & $0$ & Bolshoi+high $\sigma_8$ & This paper \\
L0250-High-$\Omega_{\rm m}$ & $250$  & $1024^3$ & $1.6 \times 10^{9}$  & $5.8$  & $1/42$  & $49$ & $0$ & Bolshoi+high $\Omega_{\rm m}$ & This paper \\
L0100-PL-1.0 & $100$  & $1024^3$ & $2.6 \times 10^{8}$  & $0.5$  & $1/195$  & $119$ & $2$ & Self--similar, $n=-1.0$ & This paper \\
L0100-PL-1.5 & $100$  & $1024^3$ & $2.6 \times 10^{8}$  & $0.5$  & $1/195$  & $99$ & $1$ & Self--similar, $n=-1.5$ & This paper \\
L0100-PL-2.0 & $100$  & $1024^3$ & $2.6 \times 10^{8}$  & $1.0$  & $1/98$  & $49$ & $0.5$ & Self--similar, $n=-2.0$ & This paper \\
L0100-PL-2.5 & $100$  & $1024^3$ & $2.6 \times 10^{8}$  & $1.0$  & $1/98$  & $49$ & $0$ & Self--similar, $n=-2.5$ & This paper
\enddata
\tablecomments{The $N$--body simulations used in this paper. $L$ denotes the box size in comoving units, $N^3$ the number of particles, $m_{\rm p}$ the particle mass, and $\epsilon$ the force softening length in physical units. More details on our system for choosing force resolutions are given in \citet{diemer_14_profiles}.}
\end{deluxetable*}

\begin{deluxetable*}{lcccccccccl}
\tablecaption{Cosmological Parameters
\label{table:cosmo}}
\tablewidth{0pt}
\tablehead{
\colhead{Cosmology} &
\colhead{$H_0$} &
\colhead{$\Omega_{\rm m}$} &
\colhead{$\Omega_{\rm \Lambda}$} &
\colhead{$\Omega_{\rm b}$} &
\colhead{$\Omega_{\rm k}$} &
\colhead{$\Omega_{\rm \nu}$} &
\colhead{$\sigma_8$} &
\colhead{$n_{\rm s}$} &
\colhead{$P(k)$} &
\colhead{Reference}
}
\startdata
Planck                         & $67$ & $0.32$ & $0.68$ & $0.0491$ & $0$ & $0$ & $0.834$ & $0.9624$ & CAMB & \citealt{planck_14_cosmology} \\
$WMAP$ (Bolshoi)                        & $70$ & $0.27$ & $0.73$ & $0.0469$ & $0$ & $0$ & $0.82$  & $0.95$   & CAMB &  \citealt{klypin_11_bolshoi}, \citet{komatsu_11} \\
Bolshoi+High-$\sigma_8$        & ''   & ''     & ''     & ''       & ''  & ''  & $0.9$   & ''       & Same as Bolshoi & \\
Bolshoi+High-$\Omega_{\rm m}$  & ''   & $0.4$  & $0.6$  & ''       & ''  & ''  & $0.82$  & ''       & Same as Bolshoi & \\
Self-similar                   & $70$ & $1$    & $0$    & $0$      & $0$ & $0$ & $0.82$  & ...       & $P(k) \propto k^n$ &
\enddata
\tablecomments{Cosmological parameters of the $N$-body simulations listed in Table \ref{table:sims}. The Bolshoi cosmology roughly corresponds to the $WMAP7$ cosmology of \citet{komatsu_11}. The Planck values correspond to the Planck-only best-fit values in Table 2 in \citet{planck_14_cosmology}. Some of the parameters in both the Planck and Bolshoi cosmologies are rounded off for convenience. The initial matter power spectrum for the Bolshoi and Planck cosmologies was computed using the Boltzmann code \textsc{Camb} \citep{lewis_00_camb}.}
\end{deluxetable*}

In this section we describe our cosmological simulations, halo identification, and resolution limits. We discuss the method used to measure concentrations, their distribution at fixed mass, and the resulting mean and median values. 

%--------------------------------------------------------------------------------------------
\subsection{$N$-body Simulations and Halo Finding}
\label{sec:numerical:sims}
%--------------------------------------------------------------------------------------------

To cover a large range of halo masses and cosmologies, we use a suite of dissipationless \LCDM simulations of different box sizes, resolutions, and cosmologies (Table \ref{table:sims}). Most of the simulations (L0031--L2000) were carried out in our fiducial cosmology, identical to that adopted in the Bolshoi simulation \citep{klypin_11_bolshoi} which is consistent with the $WMAP7$ cosmology \citep[][see Table \ref{table:cosmo}]{komatsu_11}. To investigate the dependence of concentrations on cosmology, we performed several simulations of a cosmological model consistent with recent constraints from the Planck satellite \citep{planck_14_cosmology}, as well as a number of test simulations in which only one particular parameter was changed from the value adopted in the fiducial cosmology. Finally, we ran a series of self-similar models with power-law matter power spectra of four different slopes (Table \ref{table:sims}). We use these power-law simulations to calibrate the dependence of concentration on the power spectrum slope. 

The initial conditions for the simulations were generated using the second-order Lagrangian perturbation theory code \textsc{2LPTic} \citep{crocce_06_2lptic}. The simulations were started at redshift $z = 49$ which has been shown to be sufficiently high to avoid transient effects \citep{crocce_06_2lptic}. The self-similar models with the shallowest slopes were started at higher redshift as power on small scales develops very early in such cosmologies. All simulations were computed using the publicly available code \textsc{Gadget2} \citep{springel_05_gadget2}. 

We used the phase--space halo finder \textsc{Rockstar} \citep{behroozi_13_rockstar} to extract all isolated halos and subhalos from the 100 snapshots of each simulation, and the \textsc{Consistent-Trees} code of \citet{behroozi_13_trees} to compute merger trees and establish subhalo relations. As usual, we only consider the concentrations of isolated halos, but do not exclude the contribution of subhalos to the density profiles of the isolated halos. A halo is deemed to be isolated if its center does not lie inside $\rvir$ of another, larger halo. We note that the virial radii for the various mass definitions were derived using only gravitationally bound particles. We have verified that the difference between bound masses and those including all particles is negligible for the vast majority of host halos we consider in this study.

%--------------------------------------------------------------------------------------------
\subsection{Resolution Limits}
\label{sec:numerical:limits}
%--------------------------------------------------------------------------------------------

In order to measure concentration, we need to measure the scale radius, $\rs$ (Section \ref{sec:numerical:definition}). The scale radius probes the interior of halos, and is thus susceptible to resolution effects \citep{moore_98, klypin_01}. In particular, $\rs$ needs to be resolved by a sufficient number of force resolution lengths and particles. A fixed minimum number of particles inside the virial radius does not guarantee a particular force resolution of the scale radius, because concentration depends on mass and redshift, and, more importantly, because the radius corresponding to a particular halo mass decreases with redshift due to the increasing reference density (Equations (\ref{eq:mdelta_m}) and (\ref{eq:mdelta_c})). Thus, we demand a minimum halo mass at each redshift and for each simulation that fulfills, on average, the following three criteria.
\begin{enumerate}
\item There must be at least $1000$ particles inside $\rtoc$. This requirement ensures that all $50$ radial bins used to construct the density profile are reasonably sampled.
\item There must be at least $200$ particles within $\rs$, following \citet{klypin_01} who found that the density profiles of halos converge to better than 10\% at radii enclosing at least $200$ particles.
\item $\rs$ must be at least six times the force softening length, $\epsilon$. Various authors have observed that the density profile of halos is only reliable at radii greater than 4--5 $\epsilon$ \citep{moore_98, klypin_01}. These authors used softenings that corresponded to the Newtonian force at only 2--3 softening lengths, while the \textsc{Gadget2} code used in this study employs a spline softening that reaches the Newtonian level at a smaller distance. The requirement of $r > 6\epsilon$ is thus relatively conservative. We adopt it to account for the fact that the scatter in concentrations at fixed mass is quite large, meaning that some halos have much smaller scale radii than an average halo of the same mass.
\end{enumerate}
We must not compute these criteria for each halo separately, as that would, at fixed mass, preferentially select halos of low concentration and thus bias our measurement. Instead, we compute the {\it average} scale radius at a given mass and redshift using the model of \citet{zhao_09_mah}. For the self-similar cosmologies, we find that this model does not reproduce our simulation results and thus use a fit to our own results instead. When only selecting halos that match the requirements stated above, we find that the \cmr has converged to within the statistical error. For example, making all of the requirements stricter by a factor of two does not change the mean and median relations at any redshift by more than the statistical uncertainty.

Our resolution limits are deliberately stringent because the large scatter in the \cmr means that individual halos may have smaller scale radii, and thus fewer particles and force resolution elements within $\rs$. Generally, the third criterion (the number of force softening lengths inside $\rs$) dominates the minimum mass requirement, particularly at high redshift. For example, in the L0250 simulation, the minimum halo mass is twice greater at $z = 2$ than at $z = 0$, and about $15$ times greater at $z = 6$. To appreciate the importance of our conservative cuts, let us consider the force resolution we would achieve if we only enforced the $1000$ particle limit; in this case, the scale radii of the lowest allowed mass in the L0250 simulation would, on average, be resolved by $\sim 4 \epsilon$ at $z = 0$, $\sim 3.5 \epsilon$ at $z = 2$, and $\sim 1.5 \epsilon$ at $z = 6$, leading to severe resolution effects in the concentration measurement at high redshifts \citep{moore_98, klypin_01}. 

Such resolution effects can be quantified by comparing the concentrations in simulations with very different resolutions. For example, we compare our results to the \cmr of halos in the Bolshoi simulation \citep{klypin_11_bolshoi}, measured with the same halo finder as in this paper. While the halos from our set of simulation boxes are sampled by a relatively uniform number of particles across a large range of masses, the largest halos in the Bolshoi simulation contain about $10^4$ times more particles than the smallest halos for which concentrations were measured. A high mass resolution leads to higher concentrations (see discussion in Section \ref{sec:results:model}), meaning that the Bolshoi concentrations are about $10\%$ higher than ours at high masses, and about $10\%$ lower at low masses, leading to a shallower \cmr \citep{klypin_11_bolshoi}. These results imply that the \cmr measured in simulations always tends to be biased low due to the finite mass resolution.

After we apply the resolution cuts to the halo samples from each simulation, those samples are combined into one sample per redshift. For our fiducial cosmology, this overall sample contains $\sim 86,000$ halos at $z = 0$ and $\sim 3000$ halos at $z = 6$. We do not exclude unrelaxed objects or halos that contain a large amount of substructure, although such halos are often discarded in studies of halo structure \citep[e.g.,][]{munozcuartas_11, ludlow_14_cm, dutton_14}. We choose not to exclude unrelaxed halos for several reasons. First, such a cut leads to a \cmr that is biased high because low concentrations are typical of halos in the rapid mass accretion stage which are more likely to be removed by a relaxation cut \citep[e.g.,][see also the discussion in Section \ref{sec:discussion:comp2}]{neto_07, maccio_08, bhattacharya_13}. Second, an equivalent exclusion cannot easily be performed in observations \citep[see, e.g.,][]{meneghetti_14}. Third, the fraction of excluded halos is likely a function of peak height and redshift, potentially introducing a non-universality in the \cnur. Finally, we find that the density profiles of halos even in the most active mass accretion regime are, on average, quite regular. Even though they are not as accurately fit by the NFW form as those of slowly accreting halos, their best-fit scale radii reflect real features of the profiles \citep{diemer_14_profiles}.

%--------------------------------------------------------------------------------------------
\subsection{Measuring Mean and Median Concentrations}
\label{sec:numerical:definition}
%--------------------------------------------------------------------------------------------

In this study we use the concentrations estimated by the \textsc{Rockstar} halo finder, the same software that constructs our halo catalogs and merger trees. \textsc{Rockstar} finds the scale radius of a halo by fitting the NFW profile, 
\begin{equation}
\label{eq:nfw}
\rho_{\rm NFW} = \frac{\rho_{\rm s}}{(r/\rs)(1+r/\rs)^2}
\end{equation}
to the halo's spherical mass distribution. This fit is performed using only the bound particles within $\rvir$. These particles are split into $50$ radial bins, equally spaced in enclosed mass \citep{behroozi_13_rockstar}. The contributions of the fitted NFW profile to the same mass bins are computed until the solution with the minimum bin-to-bin mass variance has been found. All bins receive equal weights, except bins at radii smaller than three force resolution softenings from the halo center, which are down-weighted by a factor of $10$. 

\begin{figure}
\centering
\includegraphics[trim = 2mm 18mm 2mm 1mm, clip, scale=0.68]{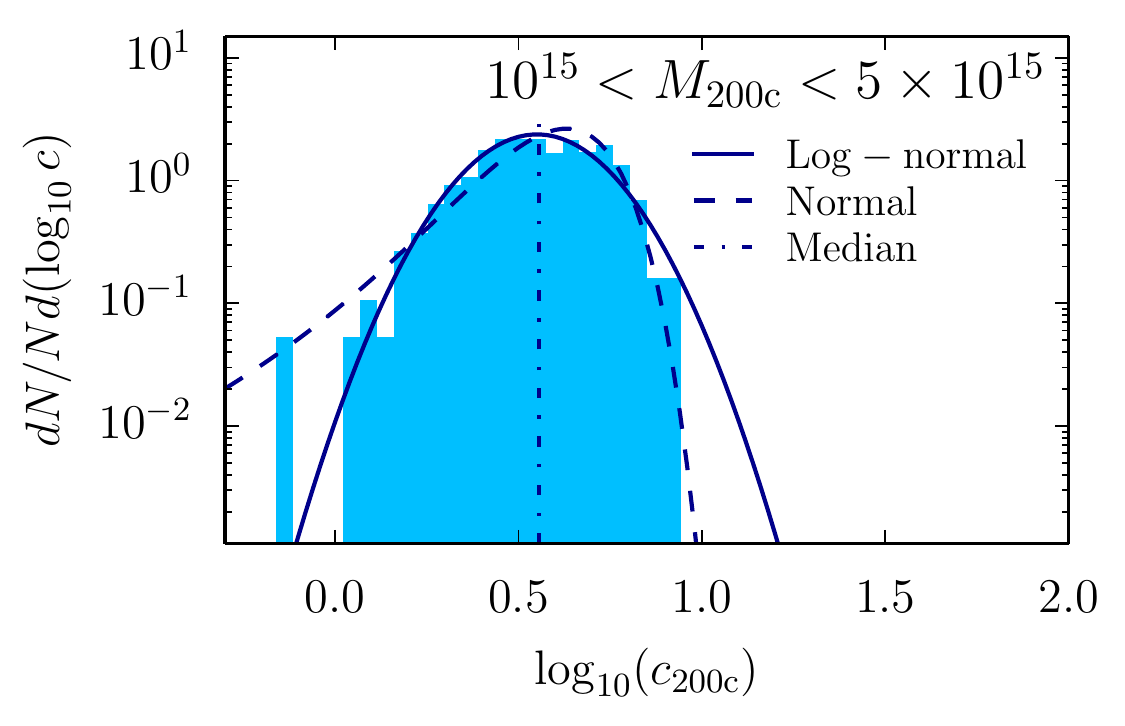}
\includegraphics[trim = 2mm 18mm 2mm 1mm, clip, scale=0.68]{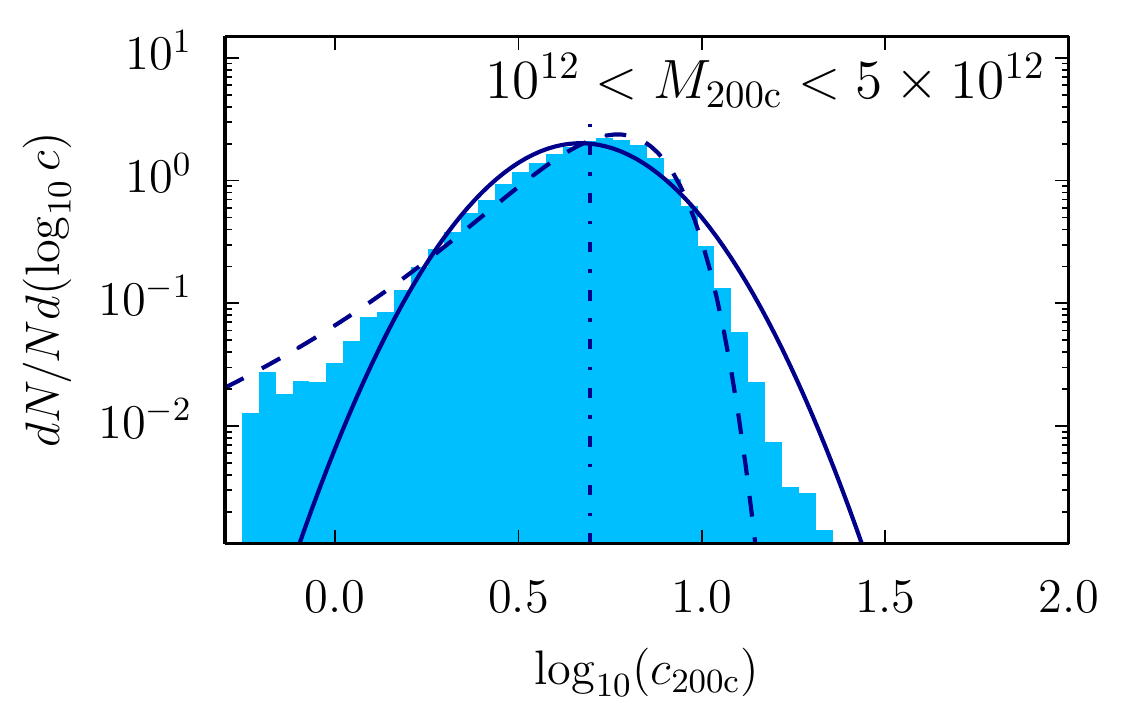}
\includegraphics[trim = 2mm 2mm 2mm 1mm, clip, scale=0.68]{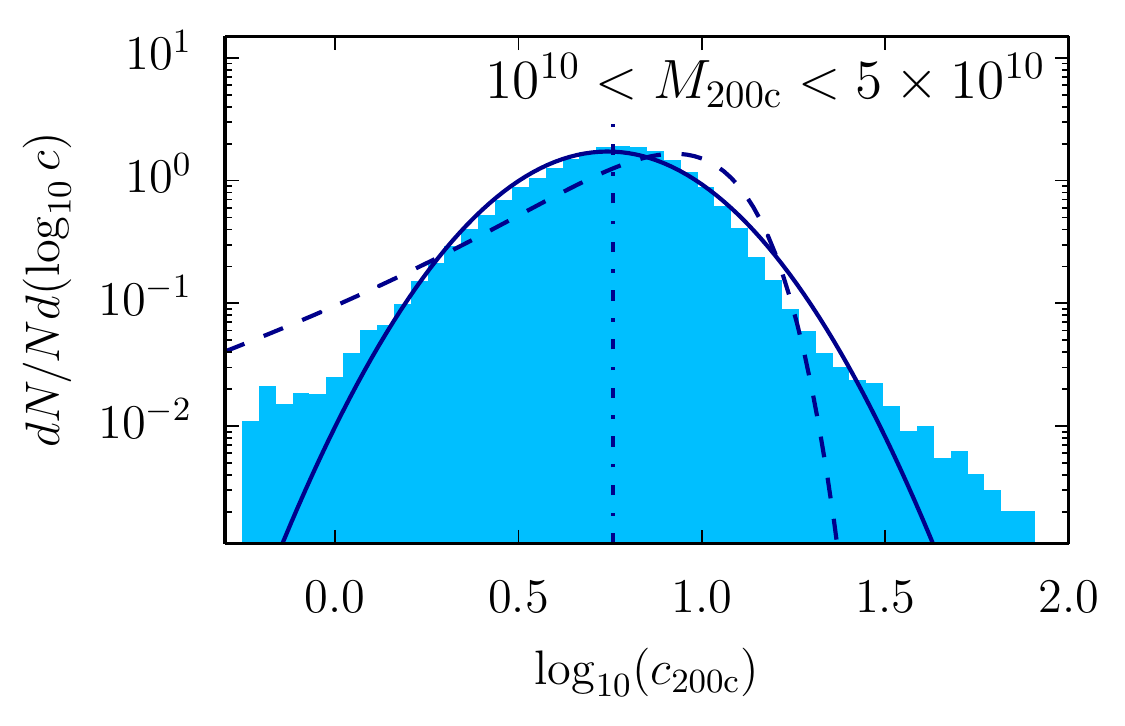}
\caption{Distribution of concentration values at high, intermediate, and low masses, at $z = 0$, and for our fiducial cosmology. Neither the normal (Gaussian) nor log-normal distributions can reproduce the measured distribution of $\ctoc$ in all three bins. The mean concentrations are not shown as they are hard to distinguish from the median given the large range of $c$ shown.}
\label{fig:distribution}
\end{figure}

While the normalization of the NFW profile, $\rho_{\rm s}$, is robust with respect to the fitting procedure, the best-fit $\rs$ can depend on technical details such as the number of bins, the radial range used in the fit, the merit function that is minimized, or the weights given to the different radial bins. These details have a  particularly large impact if the NFW profile is not a good fit to the halo profile \citep{meneghetti_13}. \citet{dooley_14} found that the mean concentrations computed by \textsc{Rockstar} are, on average, 12\% higher than concentrations found using the \textsc{Subfind} halo finder \citep{springel_01_subfind}, and that the slope of the \textsc{Rockstar} \cmr is significantly steeper. 

These conclusions, however, were derived for the mean rather than the median \cmr. The latter suffers less from extremely low or high concentration values which are often the result of poor fits, and are thus particularly dependent on the fitting algorithm used. For example, Figure \ref{fig:distribution} shows the distribution of $\ctoc$ in three mass bins at $z = 0$. The distribution of concentrations has tails at both low and high values of $c$ which are not well described by the log-normal \citep[e.g.,][]{jing_00_profiles2, bullock_01_halo_profiles, neto_07} or Gaussian \citep{reed_11, bhattacharya_13} distributions in all three mass bins. Thus, it is not clear whether the linear or logarithmic mean are a better description of the sample mean. In the high and low-mass bins, the linear mean is $6\%$ and $12\%$ larger than the median, respectively. In the intermediate-mass  bin, however, the linear mean and the median more or less coincide, while the logarithmic mean is about $6\%$ lower than the median. Thus, there is no clear preference for computing either the linear or logarithmic mean. The median, however, is much less sensitive to outliers and independent of whether the data is binned in linear or logarithmic space. Thus, we consider the median \cmr throughout this paper, unless otherwise stated. As both the mean and median concentration are of interest, we give the results of our best-fit model for both. 

We bin all masses and peak heights in log space, regardless of whether the plots are in linear or log space. The \cnu and \cm relations are binned separately, rather than translating the binned data from mass to peak height or vice versa. 

%--------------------------------------------------------------------------------------------
\section{Results}
\label{sec:results}
%--------------------------------------------------------------------------------------------

\begin{figure*}
\centering
\includegraphics[trim = 2mm 0mm 2mm 1mm, clip, scale=0.65]{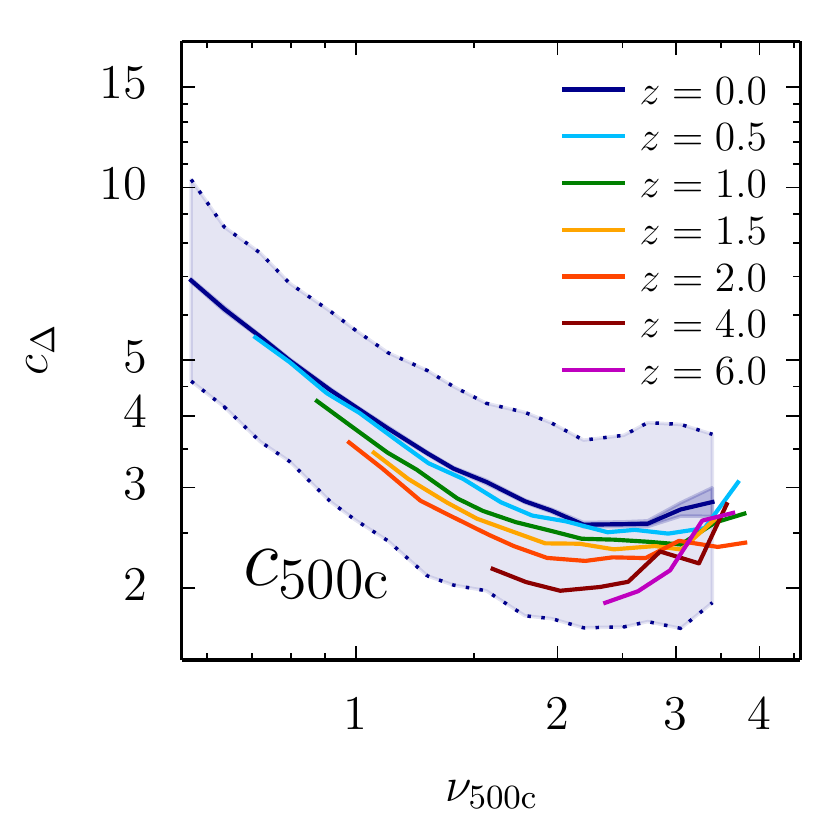}
\includegraphics[trim = 17mm 0mm 2mm 1mm, clip, scale=0.65]{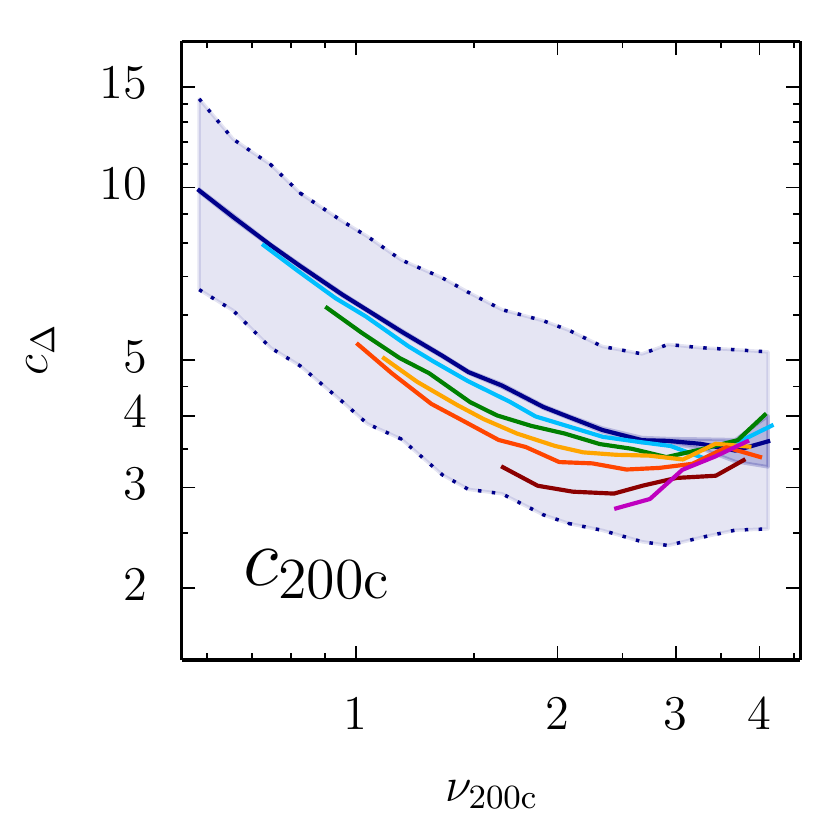}
\includegraphics[trim = 17mm 0mm 2mm 1mm, clip, scale=0.65]{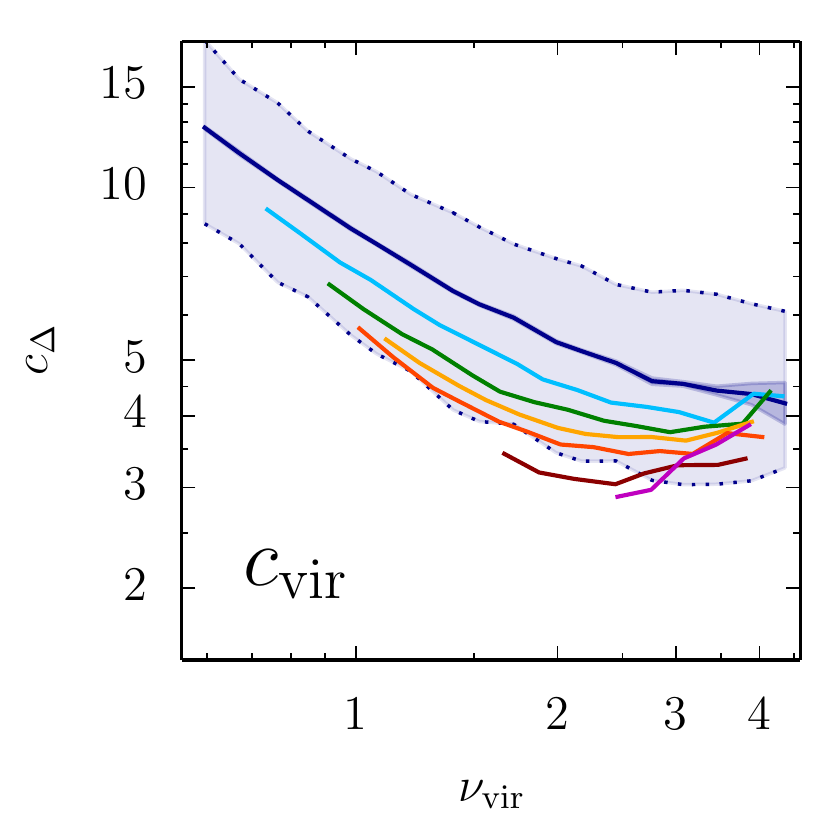}
\includegraphics[trim = 17mm 0mm 2mm 1mm, clip, scale=0.65]{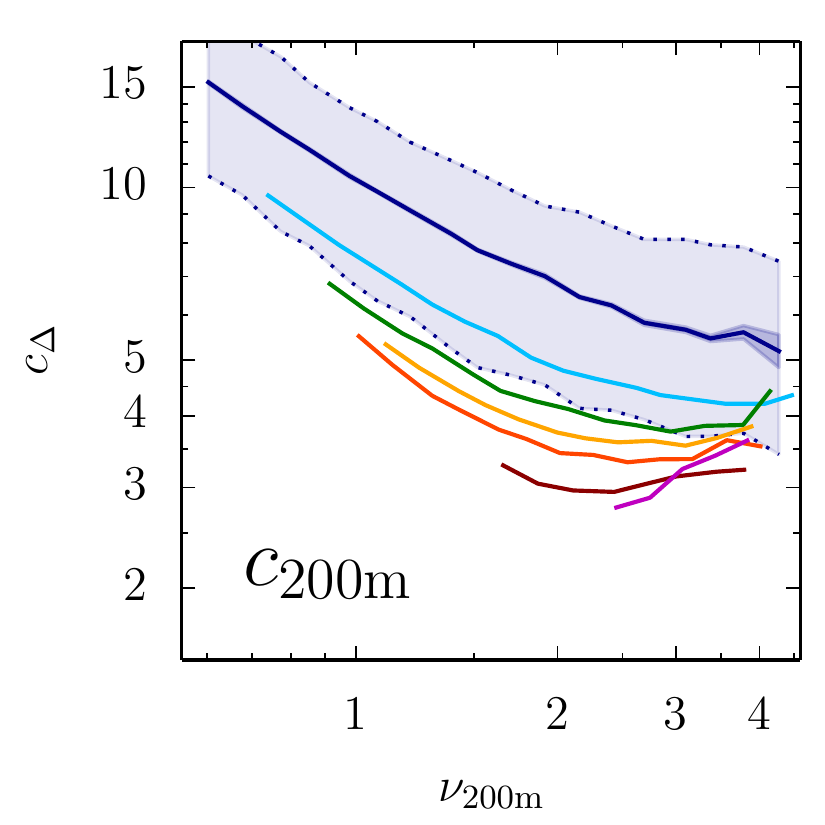}
\caption{Redshift evolution of the median concentration--peak height relation for different mass definitions. The relation in units of $\ctoc$ is significantly more universal with redshift than those in $\cvir$ and $\ctom$. However, using higher overdensities, e.g., $\cfoc$, does not improve the universality further. For consistency, $\ctoc$ is plotted against $\nutoc$ etc., but the changes in $\nu$ due to the different mass definitions are very small. The shaded area around the $z = 0$ relation indicates the 68\% scatter, the dark shaded area shows the statistical uncertainty of the median. The scatter around the relations is about $0.16$ dex at all redshifts, masses, and for all mass definitions. It is only shown at $z = 0$ to avoid overcrowding, and omitted in the rest of the plots in this paper.}
\label{fig:cm_mdefs}
\end{figure*}

In this section we present a universal, physically motivated model of halo concentrations. As a first step, we identify the optimal halo radius definition for such a model.

%--------------------------------------------------------------------------------------------
\subsection{Universality and Mass Definition}
\label{sec:results:universality}
%--------------------------------------------------------------------------------------------

As we discussed in Section \ref{sec:intro}, we can generally expect that halo structure, and thus concentration, is a universal function of the shape of the initial density peak. Indeed, numerical studies have  demonstrated that concentration is almost universal as a function of redshift at a fixed peak height, but not quite \citep{zhao_09_mah, prada_12, ludlow_14_cm, dutton_14}.

Given that there are many commonly used definitions of the ``virial'' radius, and thus many definitions of concentration (e.g., $\cfoc$, $\ctoc$, $\cvir$, and $\ctom$; see Section \ref{sec:numerical:definitions}), we ask the question whether concentrations are most universal for a certain definition. For example, in \citet{diemer_14_profiles} we showed that the density profiles of halo samples of the same peak height, but at different redshifts, are most universal at small radii ($r \lsim \rtoc$) when radii are rescaled by $\rtoc$, whereas they are most universal in units of $\rtom$ at large radii. As concentration is a property of the inner halo profile, it stands to reason that the \cnur should be most universal across redshift when $\ctoc$ is used. Furthermore, \citet{dutton_14} hint at a stronger redshift evolution of definitions other than $\ctoc$, and \citet{bhattacharya_13} showed that there is a difference in the evolution of $\cvir$ and $\ctoc$ at high masses. 

Figure \ref{fig:cm_mdefs} shows the \cnurs for the $\cfoc$, $\ctoc$, $\cvir$, and $\ctom$ definitions up to $z = 6$. It is clear that the choice of definition has a large impact on the degree of non-universality in the relations: while the \ctocnur comes closest to universality, the $\cvir$-$\nu$ and $\ctom$-$\nu$ relations exhibit a much larger evolution at low redshifts. For example, at $\nu  = 1$, $\ctoc$ barely changes between $z = 1$ and $z = 0$, but the corresponding $\ctom$ evolves from $\sim 6$ at $z = 1$ to $\sim 11$ at $z = 0$. The differences appear at low $z$ because that is the epoch when $\Omega_{\rm m}$ drops below unity and dark energy starts to dominate, which leads to a different evolution of $\rhom$ and $\rhoc$. The universality does not further improve over $\ctoc$ when using definitions with a higher overdensity threshold, such as $\cfoc$ (left panel of Figure \ref{fig:cm_mdefs}).

The results in Figure \ref{fig:cm_mdefs} clearly demonstrate that $\ctoc$ is preferable when devising a universal model for the {\cnur}. Many previous works on the \cmr have, in fact, used $\ctoc$ as their measure of concentration \citep{navarro_96, navarro_97_nfw, jing_00_profiles2, neto_07, duffy_08, gao_08, prada_12, ludlow_14_cm, dutton_14}, but some authors used $\cvir$ \citep{bullock_01_halo_profiles, wechsler_02_halo_assembly, zhao_09_mah, klypin_11_bolshoi, munozcuartas_11} or even $\ctom$ \citep{dolag_04}. Of course, the universality of the \cnur may not be the only consideration when choosing a concentration definition. Different definitions also lead to different redshift evolutions of the concentration of individual halos, and thus different shapes of the \cmr (see Appendix \ref{sec:discussion:evolution}). Given the results presented in Figure \ref{fig:cm_mdefs}, we focus on $\ctoc$ for the remainder of this paper.

Although the \ctocnur is more universal than definitions using lower overdensities, it still exhibits sizeable deviations from universality of up to $25\%$ around $\nu \approx 2$, significantly larger than the statistical uncertainty on the relations. We have checked that these deviations are not due to resolution effects by comparing the \ctocnur in simulations with different resolution. Similar results were also recently found by \citet{dutton_14}. The non--universality of the \cnur indicates that there is at least one additional parameter besides peak height that influences concentration.

%--------------------------------------------------------------------------------------------
\subsection{Analytical Model for Halo Concentrations}
\label{sec:results:model}
%--------------------------------------------------------------------------------------------

Let us consider the parameters that might control halo concentrations and their evolution. As shown in previous studies, the mass dependence of the concentration of halos is a consequence of their MAH \citep{wechsler_02_halo_assembly, zhao_03_mah, zhao_09_mah, ludlow_13_mah}, which, in turn, is determined by the parameters of the background cosmological model. These cosmological parameters control both the growth rate of the initial fluctuations as a function of time and the linear matter power spectrum, $P(k)$. The latter determines the statistical properties of the peaks in the initial density field, such as their peak height as a function of scale, and their curvature. As discussed earlier, expressing halo masses as peak heights should, in principle, account for the dependence of concentration on cosmological parameters. 

\begin{figure*}
\centering
\includegraphics[trim = 6mm 4mm 0mm 1mm, clip, scale=0.73]{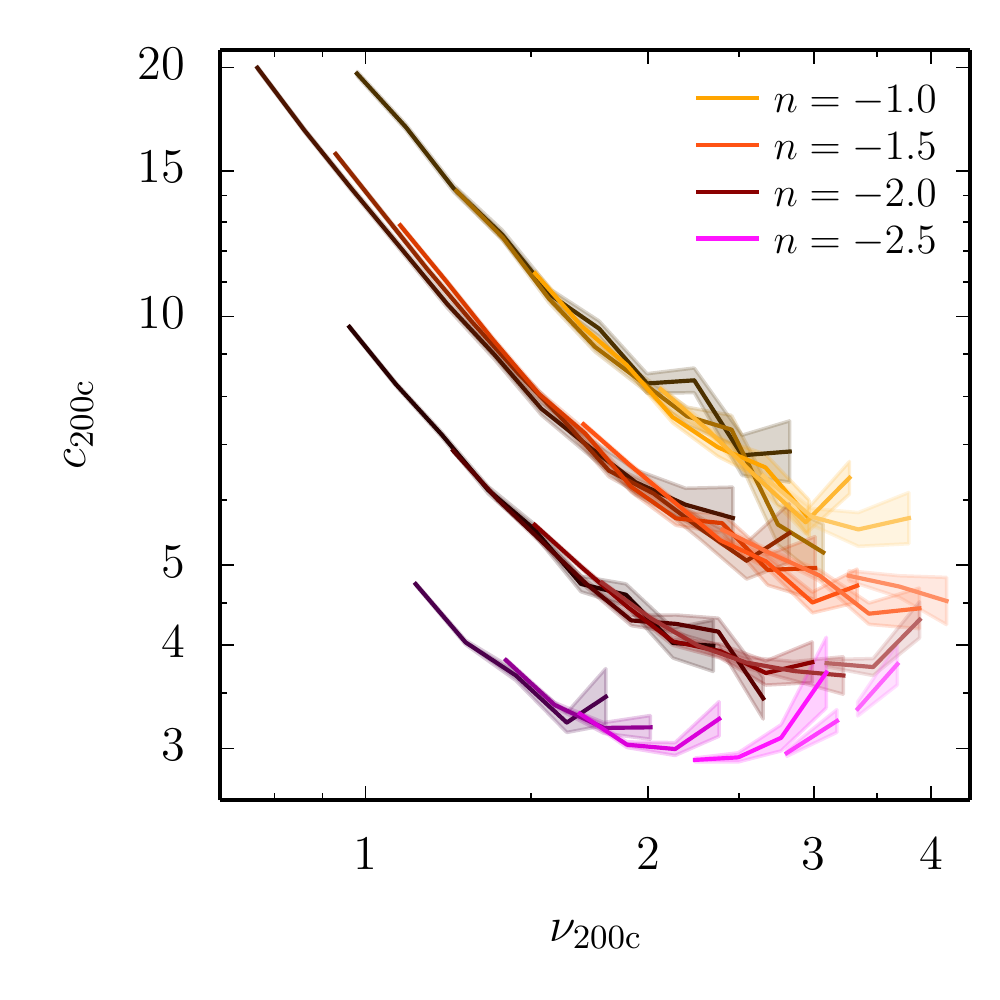}
\includegraphics[trim = 2mm 4mm 0mm 1mm, clip, scale=0.73]{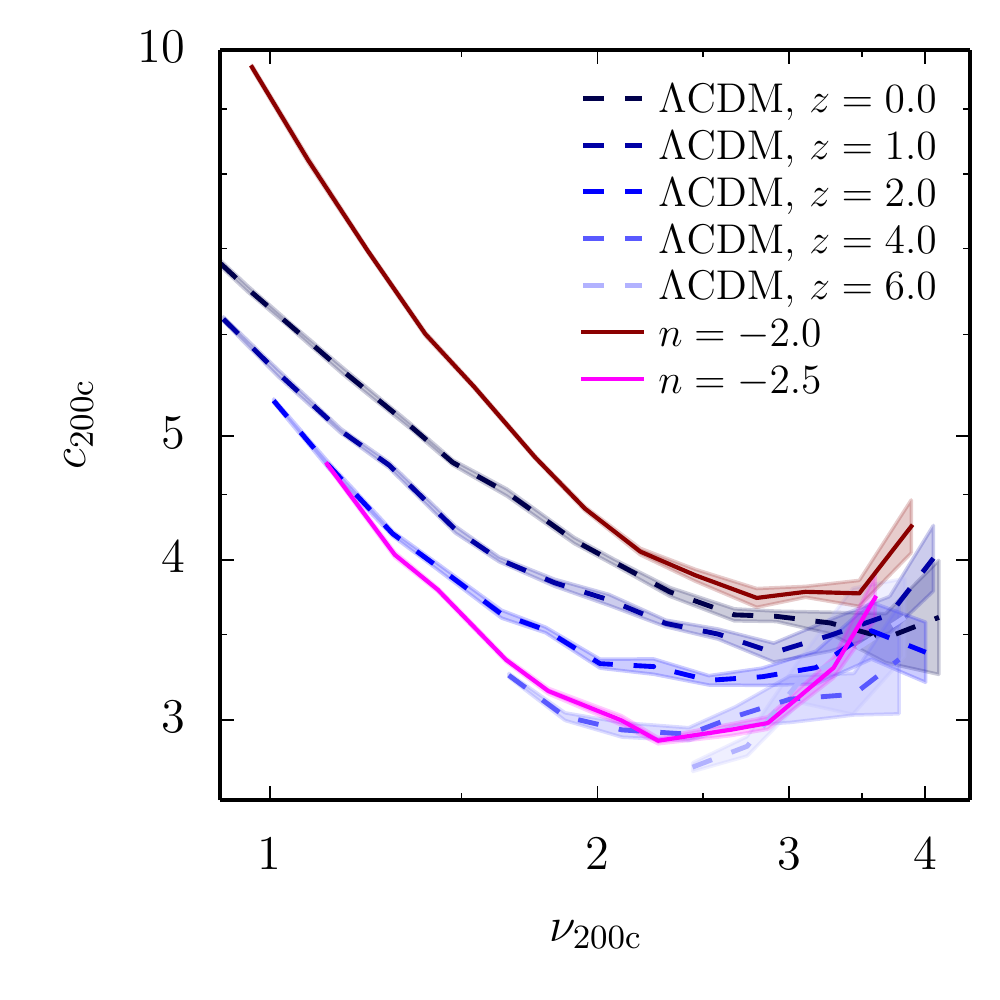}
\caption{Median \cnurs for self-similar (left panel) and $\Lambda$CDM (right panel) cosmologies. The shaded areas show the statistical uncertainty around the median relations. Left panel: concentrations in the four self-similar cosmologies (Table \ref{table:sims}). The colors correspond to the four different slopes $n$ ($-1$, $-1.5$, $-2$, and $-2.5$), while the shading of the lines indicates redshift, with darker lines corresponding to lower redshifts.  The respective redshifts are ($2$, $3$, $4$, $6$, $8$) for $n = -1$, ($1$, $1.5$, $2$, $4$, $6$, $8$) for $n = -1.5$, ($0.5$, $1$, $1.5$, $2$, $4$) for $n = -2$, and ($0$, $0.25$, $0.5$, $1$, $1.5$, $2$) for $n = -2.5$. As expected, the \cnur does not evolve with redshift in power-law cosmologies.  Right panel: comparison of the power-law models with $n = -2$ and $n = -2.5$ to various redshifts in our fiducial $\Lambda$CDM cosmology (dashed blue lines, darker color indicating lower redshift). The various redshifts in the power-law cosmologies have been combined into one relation per simulation (solid lines). The comparison demonstrates that the changing shape of the $\Lambda$CDM \cnur with redshift is likely related to the changing local slope of the power spectrum at a fixed $\nu$.}
\label{fig:powerlaw}
\end{figure*}

\begin{figure*}
\centering
\includegraphics[trim = 1mm 7mm 2mm 5mm, clip, scale=0.64]{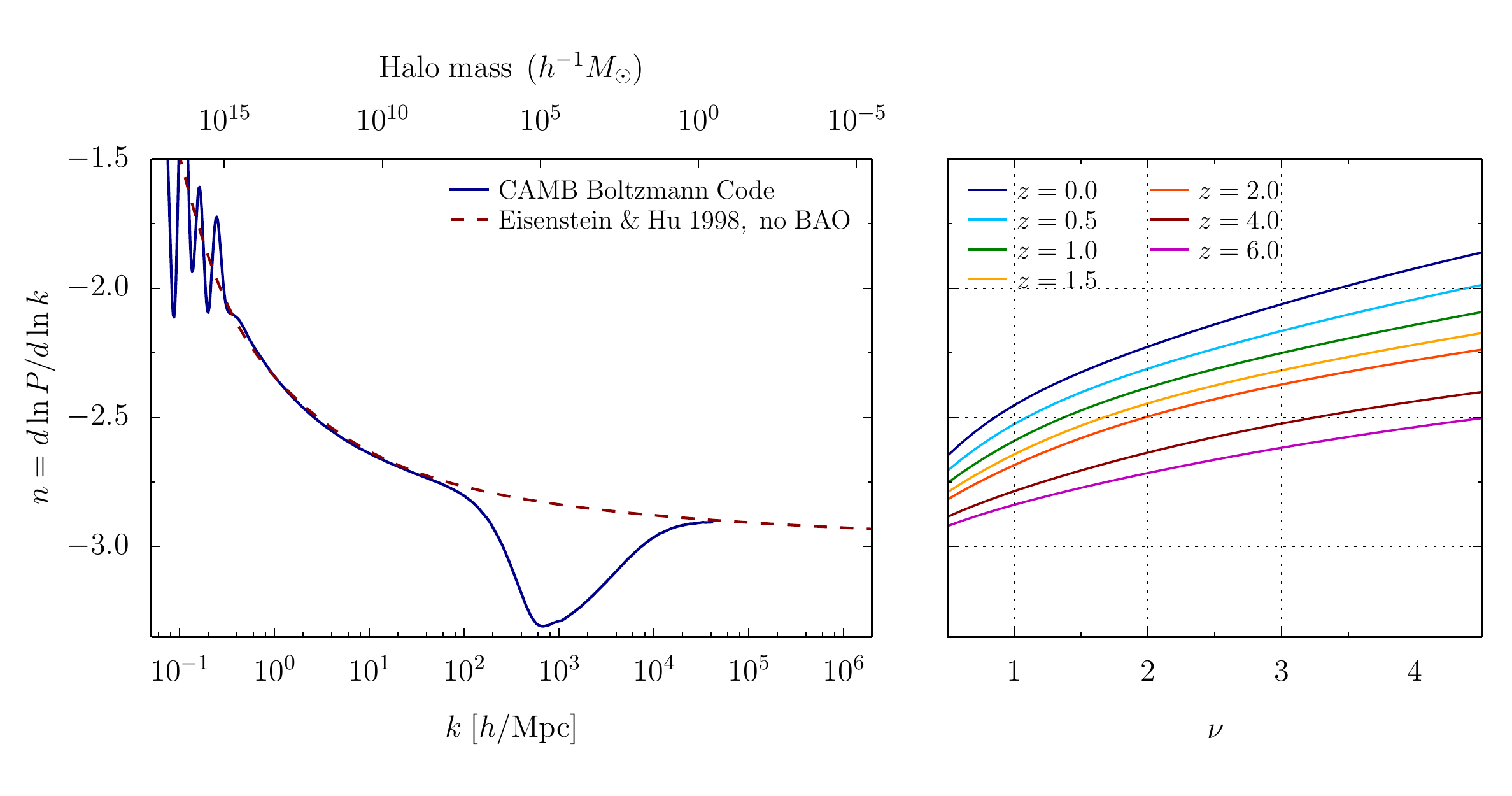}
\caption{Logarithmic slope of the linear matter power spectrum, $n$, evaluated at different scales, peak heights, and redshifts. Left panel: the slope of $P(k)$ computed from the power spectrum produced by the \textsc{Camb} Boltzmann code \citep[][solid blue line]{lewis_00_camb}, as well as from the $P(k)$ approximation of \citet[][dashed red line]{eisenstein_98} without baryon acoustic oscillations to avoid oscillatory behavior in $n$ for the largest halos. The top axis shows the halo mass corresponding to the Lagrangian volume of radius $R = 2\pi / k$. This scale gives a rough indication of what part of the power spectrum is most important for the formation of halos of a given mass. The decrease in power in the \textsc{Camb} spectrum around $k \approx 100 h/{\rm Mpc}$ is caused by the pressure of baryons which is not modeled in the \citet{eisenstein_98} approximation. We note, however, that the Nyquist frequency of our smallest simulation box is $k_{\rm N} = 103 h/{\rm Mpc}$ so that these small scales are barely resolved. Right panel: the slope at $k_{\rm R}(\nu, z) = \kappa\, 2\pi / R$ (Equation (\ref{eq:kr})), evaluated for halos of different peak heights at different redshifts. At high $z$, the masses and radii corresponding to a fixed $\nu$ are smaller, leading to steeper slopes.}
\label{fig:pkslope}
\end{figure*}

However, Figure \ref{fig:cm_mdefs} demonstrates that there is a residual dependence on at least one additional parameter. It is well known that halo concentrations depend on the power spectrum slope in self-similar models \citep[][see also Figure \ref{fig:powerlaw}]{navarro_97_nfw, eke_01_concentrations, reed_05, zhao_09_mah}. A natural candidate for the additional parameter is thus the local slope of the power spectrum,
\begin{equation}
n(k) \equiv \frac{d \ln P(k)}{d \ln k} \,,
\end{equation}
which for CDM cosmologies changes as a function of physical scale $k$. A fixed $\nu$ corresponds to different masses at different redshifts, and thus corresponds to different values of $n(k)$. 

Physically, $n$ can affect concentrations in two distinct ways. First, it determines the steepness of the mass function of halos that merge with a given halo at different times \citep{lacey_93_collapse}. It has been shown that the amount of substructure in the accreted matter influences the concentration of a halo \citep{moore_99}. While the matter accreted smoothly or in low--mass halos is distributed to radii determined by its energy and angular momentum, massive subhalos can lose angular momentum due to dynamical friction and sink to the center of the accreting halo, increasing its concentration \citep{chandrasekhar_43, boylankolchin_08}. The magnitude of this effect has been subject to some debate \citep{moore_99, huss_99, colin_08}. Although the effect is relatively small, it is potentially sufficient to explain the modest deviations from universality observed in the \cnur. At the same time, in the regime of a steep power spectrum, halos of different masses collapse closer in time and major mergers will thus be more frequent. This will result in a large fraction of unrelaxed halos, which is known to affect the normalization and slope of the \cnur \citep{ludlow_12}.

Second, $n$ can affect concentrations via its effect on the shape of the initial density peaks. \citet{dalal_10} demonstrated that the MAH of a halo and its density profile are tightly connected to the density profile of the corresponding initial peak. Thus, one might expect that some parameter describing the peak shape should be used in our model. To this end, we have explored the average curvature of peaks of a given height \citep{bardeen_86, dalal_10} as a possible second parameter affecting concentrations at fixed $\nu$. We found, however, that peak curvature cannot by itself explain deviations from universality observed in the \cnur (see Appendix \ref{sec:discussion:curvature} for a detailed discussion). 

\begin{figure*}[ht!]
\centering
\includegraphics[trim = 2mm 3mm 0mm 2mm, clip, scale=0.75]{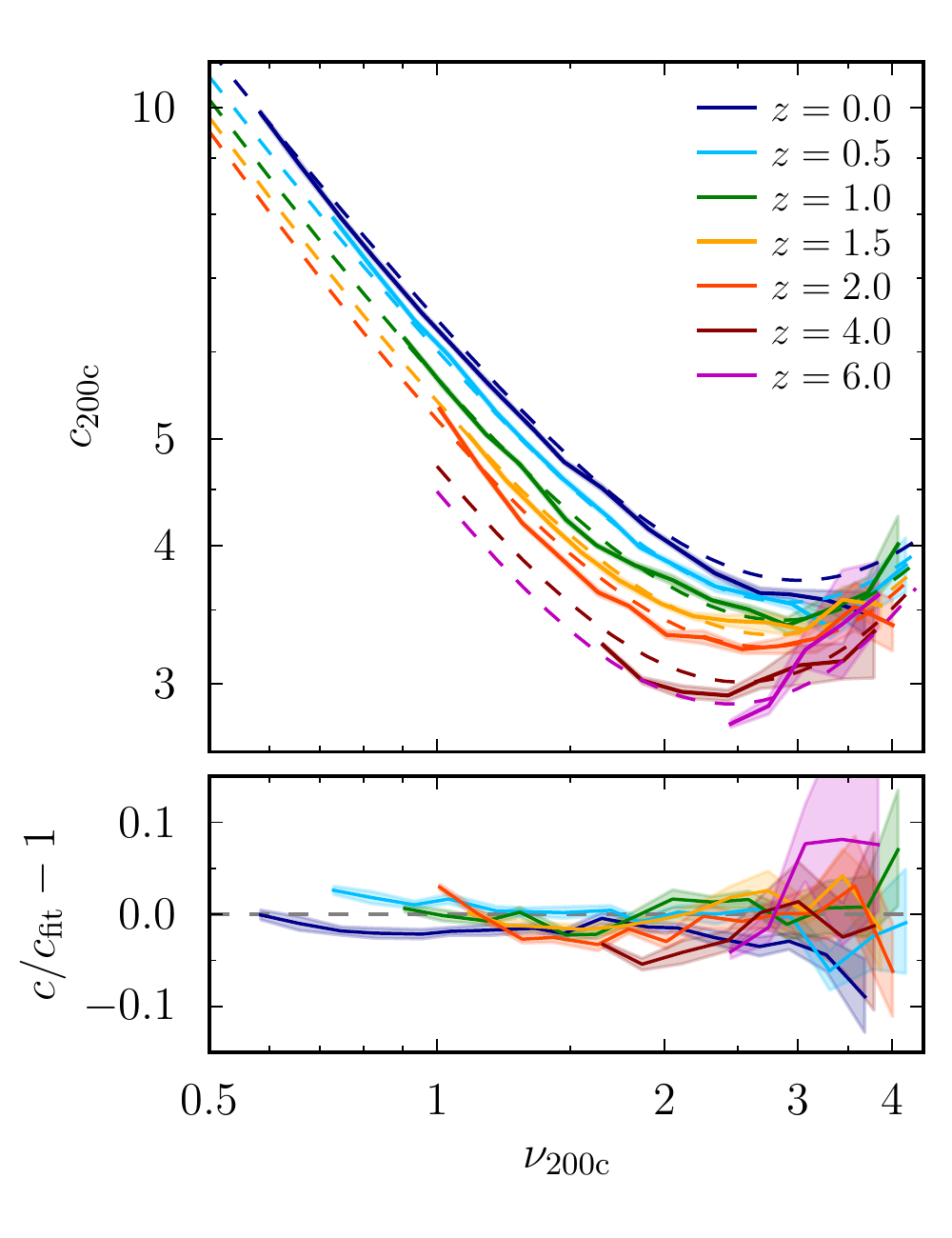}
\includegraphics[trim = 20mm 3mm 0mm 2mm, clip, scale=0.75]{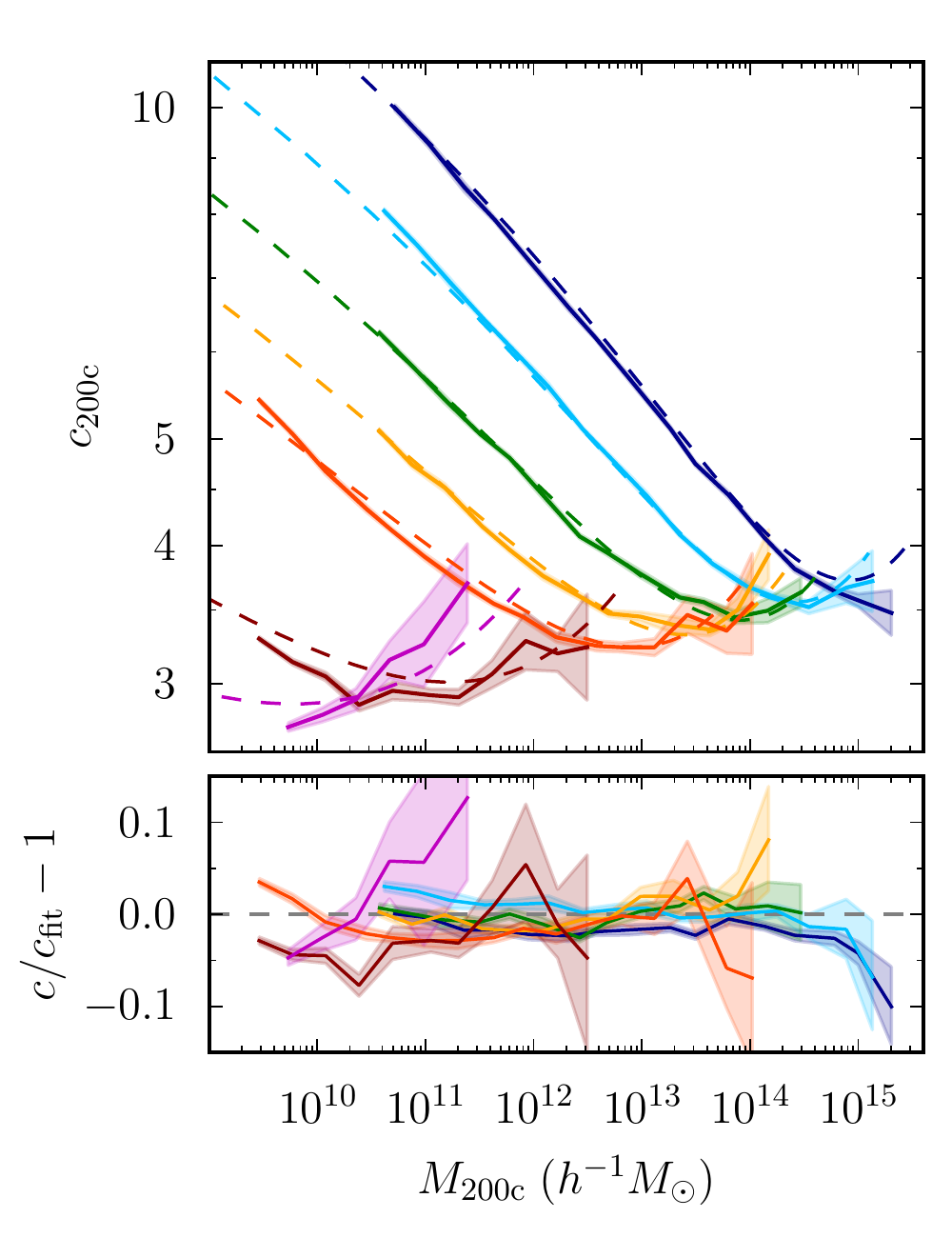}
\caption{Comparison of our model with simulation data for the fiducial $\Lambda$CDM cosmology. The dashed lines show the median \cnu (left) and \cm (right) relations predicted by our $c(\nu,n)$ model, whereas the solid lines and shaded areas show the median concentrations of simulated halos and their statistical uncertainties. Our model fits the measured relations to better than $\sim 5\%$ at those $\nu$ and $z$ where $c$ is measured reliably. The steepening of the slope of the local power spectrum at higher $z$ explains the decrease of the minimum concentrations and the more pronounced upturn in the \cnur at high $\nu$.}
\label{fig:match_bolshoi}
\end{figure*}

We have chosen $n$ as a second parameter for our model because it likely captures a combination of the substructure and peak shape effects. While we do not have a solid physical model predicting this overall effect, we calibrate it using simulations of power-law cosmologies (Table \ref{table:sims}). In such models, the \cnur is expected to be universal across redshifts for a given cosmology, but to depend on $n$, the only input parameter of the model. The left panel of Figure \ref{fig:powerlaw} shows the \cnu relations in our self-similar simulations with $n = -1$, $-1.5$, $-2$, and $-2.5$. For each of these simulations, multiple redshifts are shown in different shadings. As expected, the relations measured at different redshifts are consistent with a single, universal \cnur\ {\it for a model with a given $n$}. However, the figure clearly shows that the overall \cnur does depend strongly on $n$ at a fixed $\nu$. To quantify this dependence, we fit the \cnurs of the self-similar models with the following double power-law function,
\begin{equation}
\label{eq:fitfunc}
\ctoc = \frac{c_{\rm min}}{2} \left[ \left( \frac{\nu}{\nu_{\rm min}} \right)^{-\alpha} + \left( \frac{\nu}{\nu_{\rm min}} \right)^\beta\right],
\end{equation}
where the concentration floor, $c_{\rm min}$, and its location, $\nu_{\rm min}$, are assumed to depend linearly on the power spectrum slope,
\begin{align}
\label{eq:fitfunc2}
c_{\rm min} &= \phi_0 + \phi_1 n \nonumber \\
\nu_{\rm min} &= \eta_0 + \eta_1 n \,,
\end{align}
while the slopes $\alpha$ and $\beta$ are fixed. This functional form matches the results of all four self-similar simulations well with only six parameters which can be determined via a least-squares fit. 

However, we are primarily interested in devising a model that works for all cosmologies, and is particularly accurate for $\Lambda$CDM. In $\Lambda$CDM models, $n$ varies over a considerably narrower range than in the self-similar models we have explored. The right panel of Figure \ref{fig:powerlaw} shows the results of the $n = -2$ and $n = -2.5$ simulations, as well as the \cnurs in our fiducial $\Lambda$CDM cosmology at different redshifts. The $\Lambda$CDM relations mostly lie between the relations for the $n = -2$ and $n = -2.5$ models, which approximately corresponds to the range of $P(k)$ slopes at the relevant scales in the $\Lambda$CDM power spectrum. 

However, before we can test whether our model in Equation (\ref{eq:fitfunc}) also applies to $\Lambda$CDM cosmologies, we need to properly define $n$ for such models. The simplest definition of $n$ is the local slope of $P(k)$ at some scale $k$. A natural scale for the ``size'' of a halo is its Lagrangian radius, 
\begin{equation}
\label{eq:lagrangian_radius}
R = \left( \frac{3 M}{4 \pi \rho_{\rm m}(z = 0)} \right)^{1/3} \,,
\end{equation}
the same expression used to convert halo mass to radius in the definition of peak height in Equation (\ref{eq:mtor}). The left panel of Figure \ref{fig:pkslope} shows the logarithmic slope of $P(k)$ for our fiducial cosmology as a function of scale, $k$. The corresponding mass scales, $k = 2\pi/R$, are indicated in the top axis. We compute $P(k)$ using the approximation of \citet{eisenstein_98}, namely the version without baryon acoustic oscillations as they introduce oscillatory behavior at the very highest halo masses. For comparison, we also show the slope of the exact power spectrum, computed by the Boltzmann code \textsc{Camb} \citep{lewis_00_camb}. 

\begin{deluxetable}{lllc}
\tablecaption{Best-fit Parameters
\label{table:params}}
\tablewidth{0pt}
\tablehead{
\colhead{Param.} &
\colhead{Value} & 
\colhead{Description} & 
\colhead{Equ.}
}
\startdata
\\
\multicolumn{4}{l}{Definition of Power Spectrum Slope} \\
\hline
\rule{0pt}{3ex} $\kappa$         & 0.69  & Location in $k$-space where slope is evaluated & \ref{eq:kr} \\
\\
\multicolumn{4}{l}{Best-fit \cnur (Median)} \\
\hline
\rule{0pt}{3ex} $\phi_0$   & $6.58$  & Normalization of concentration floor & \ref{eq:fitfunc2} \\
\rule{0pt}{0pt} $\phi_1$   & $1.37$  & Slope dependence of concentration floor & \ref{eq:fitfunc2} \\
\rule{0pt}{0pt} $\eta_0$   & $6.82$  & Normalization of $\nu$ where $c$ is minimal & \ref{eq:fitfunc2} \\
\rule{0pt}{0pt} $\eta_1$   & $1.42$  & Slope dependence of $\nu$ where $c$ is minimal & \ref{eq:fitfunc2} \\
\rule{0pt}{0pt} $-\alpha$  & $-1.12$ & Slope of \cnur at low $\nu$ & \ref{eq:fitfunc} \\
\rule{0pt}{0pt} $\beta$    & $1.69$  & Slope of \cnur at high $\nu$ & \ref{eq:fitfunc} \\
\\
\multicolumn{4}{l}{Best-fit \cnur (Mean)} \\
\hline
\rule{0pt}{3ex} $\phi_0$   & $7.14$  & Normalization of concentration floor & \ref{eq:fitfunc2} \\
\rule{0pt}{0pt} $\phi_1$   & $1.60$  & Slope dependence of concentration floor & \ref{eq:fitfunc2} \\
\rule{0pt}{0pt} $\eta_0$   & $4.10$  & Normalization of $\nu$ where $c$ is minimal & \ref{eq:fitfunc2} \\
\rule{0pt}{0pt} $\eta_1$   & $0.75$  & Slope dependence of $\nu$ where $c$ is minimal & \ref{eq:fitfunc2} \\
\rule{0pt}{0pt} $-\alpha$  & $-1.40$ & Slope of \cnur at low $\nu$ & \ref{eq:fitfunc} \\
\rule{0pt}{0pt} $\beta$    & $0.67$  & Slope of \cnur at high $\nu$ & \ref{eq:fitfunc} \\
\\
\multicolumn{4}{l}{Scatter (Independent of $M$, $z$, or Mass Definition)} \\
\hline
\rule{0pt}{3ex} $\sigma$ & 0.16 & 68\% scatter in concentration (dex) & $...$ \\
\enddata
\end{deluxetable}

The scale of the relevant effective slope will not, in general, be equal to the scale defined in Equation (\ref{eq:lagrangian_radius}) because the mass spectrum of halos with which a given halo merges is determined by the slope on scales smaller than $R$. Nevertheless, we expect the bulk of the effect on concentration to be caused by mergers with relatively massive halos, comparable in scale to the main halo itself. Hence, one can argue that the range of scales over which the effective slope should be measured is relatively small and should be close to $R$. Thus, we define the effective wavenumber,
\begin{equation}
\label{eq:kr}
k_{\rm R}(\nu, z) \equiv \kappa \frac{2 \pi}{R},
\end{equation}
which corresponds to a wavelength of $1 / \kappa$ times the Lagrangian radius of a halo, where $\kappa$ is a free parameter defining the scale of the effective slope. Through experimentation, we found that the local slope at $k_{\rm R}$ with $\kappa$ close to $1$ (see Table \ref{table:params} for the exact value) is a suitable estimate of the effective $n$ for our purposes. The right panel of Figure \ref{fig:pkslope} shows the resulting $n(\nu, z)$ for the redshifts and peak heights considered in this study. At higher $z$, a fixed $\nu$ corresponds to smaller halos, smaller scales, and thus steeper slopes. 

\begin{figure}[!t]
\centering
\includegraphics[trim = 2mm 19mm 2mm 81mm, clip, scale=0.78]{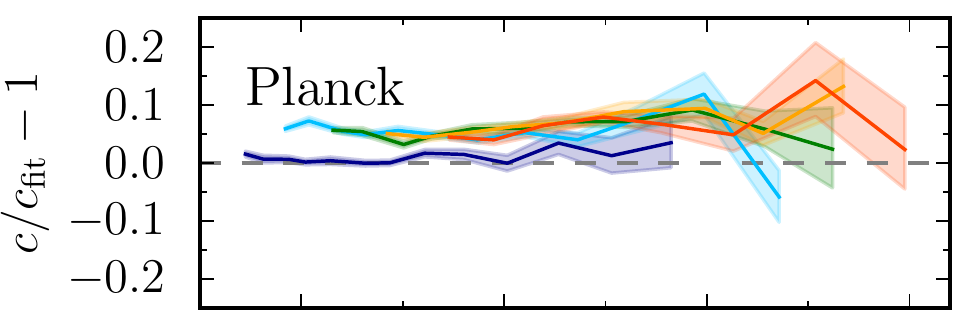}
\includegraphics[trim = 2mm 19mm 2mm 81mm, clip, scale=0.78]{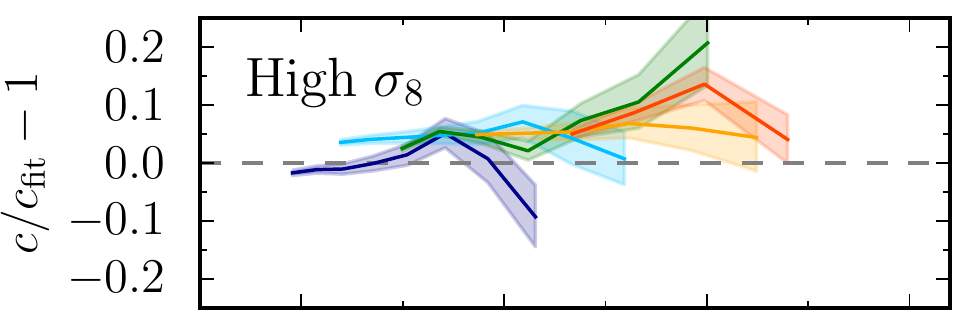}
\includegraphics[trim = 2mm 3mm 2mm 81mm, clip, scale=0.78]{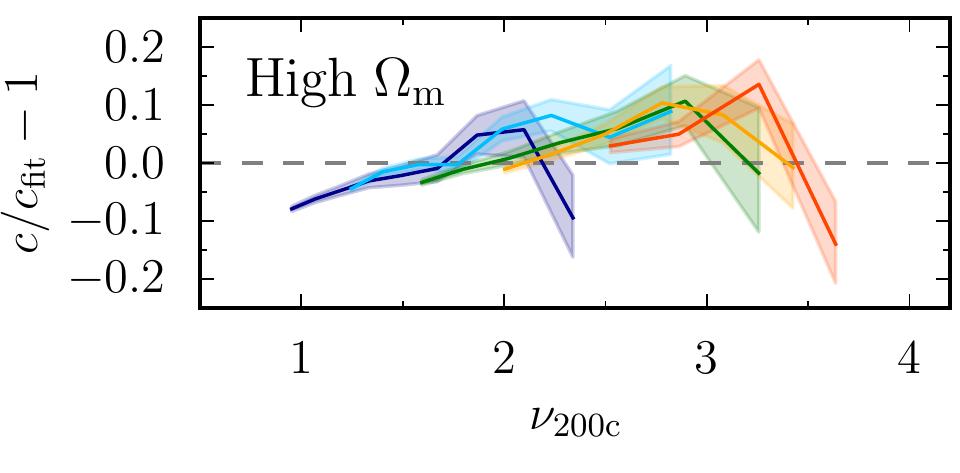}
\caption{Same as Figure \ref{fig:match_bolshoi}, but for the Planck, High-$\sigma_8$, and High-$\Omega_{\rm m}$ cosmologies. Given that the \cnu and \cm relations for these cosmologies are relatively similar to those in the Bolshoi cosmology, only the relative differences between our model and the simulation results are shown. Our model describes the relations to $\lesssim 10\%$ accuracy. See text for a detailed discussion.}
\label{fig:match_other}
\end{figure}

\begin{figure}[!t]
\centering
\includegraphics[trim = 2mm 3mm 2mm 2mm, clip, scale=0.75]{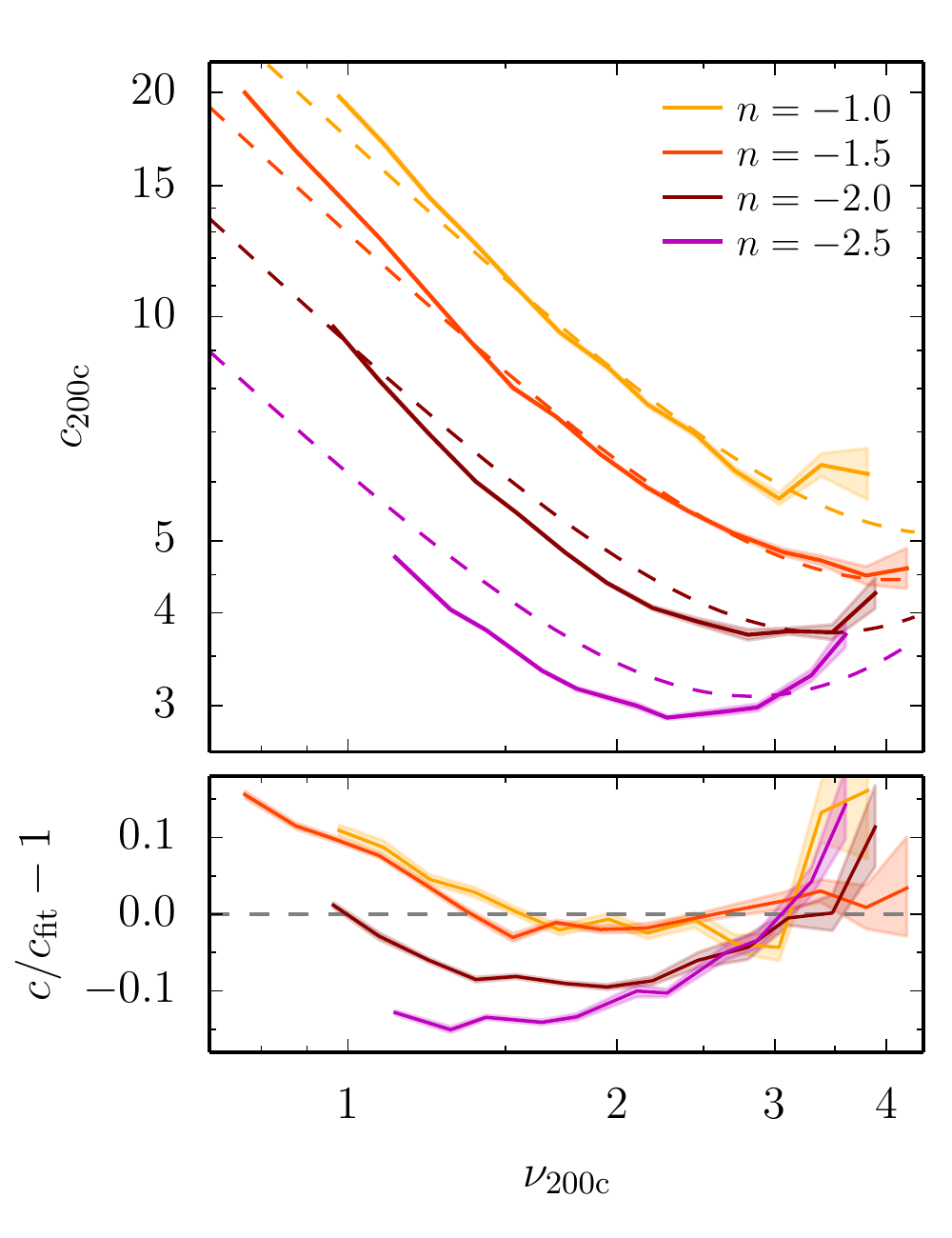}
\caption{Same as Figure \ref{fig:match_bolshoi}, but for the self-similar cosmologies. These models received a lower weight in the parameter fit as they are deemed less important than the more realistic $\Lambda$CDM cosmology, and our model thus fits them to $\sim 15\%$ accuracy rather than the $\sim 5\%$ accuracy achieved for $\Lambda$CDM.}
\label{fig:match_pl}
\end{figure}

We are now in a position to test whether the fitting model of Equation (\ref{eq:fitfunc}) works for $\Lambda$CDM as well as for the power-law cosmologies. In particular, we seek a set of best--fit values of the six parameters in Equation (\ref{eq:fitfunc}), as well as a best-fit value of $\kappa$, which lead to a good fit to the simulated \cnurs in the power-law cosmologies, our fiducial cosmology, and the Planck, High-$\sigma_8$ and High-$\Omega_{\rm m}$ cosmologies. We estimate these best-fit parameters by performing a global least-squares fit over the simulation results for all cosmologies, masses, and redshifts. We assign the $\Lambda$CDM cosmologies a weight five times larger than that of the power-law cosmologies because we want to achieve the highest accuracy for the $\Lambda$CDM models. As a result, our model matches the $\Lambda$CDM cosmologies to better than $\approx 5\%$ at all redshifts and masses where the \cnur is determined reliably, while it matches the power-law cosmologies to better than $\approx 15\%$. Detailed comparisons between our model and simulation data are shown in Figure \ref{fig:match_bolshoi} for the fiducial cosmology, Figure \ref{fig:match_other} for the Planck, High-$\sigma_8$, and High-$\Omega_{\rm m}$ cosmologies, and Figure \ref{fig:match_pl} for the self-similar cosmologies. The best-fit parameters for the mean and median \cnurs are listed in Table \ref{table:params}. Our model natively predicts $\ctoc$, and so Figures \ref{fig:match_bolshoi}--\ref{fig:match_pl} show comparisons in this mass definition. However, the $\ctoc$ predictions can be converted to other mass definitions assuming a particular form of the density profile, as discussed in Appendix \ref{sec:discussion:mdefs}.

Figure \ref{fig:match_bolshoi} shows that our model naturally captures the shape of the \cnu and \cm relations in the $\Lambda$CDM cosmology at high redshifts, namely the decreasing minimum concentration and the progressively more positively sloped relations at high $\nu$ and $M$. We note that the slopes at low and high $\nu$, $\alpha$ and $\beta$, are constant with redshift. However, because the simulations probe different ranges of $\nu$ at different redshifts, the slope of the relation appears to evolve.

The top panel of Figure \ref{fig:match_other} shows the residuals between our model and the \cnurs in the Planck cosmology. At fixed mass, the Planck concentrations are $\approx 15\%$ higher than those in our fiducial cosmology, in agreement with the results of \citet{dutton_14}. This shift is mostly caused by the higher values of $\Omega_{\rm m}$ and $\sigma_8$ \citep{dooley_14}. At $z = 0$, our model fits the Planck cosmology simulation results very well, whereas it underestimates the Planck concentrations at $z > 0$ by $\approx$ 5--10$\%$, within the statistical accuracy of our model. Similarly, our model fits the High-$\sigma_8$ and High-$\Omega_{\rm m}$ \cnurs reasonably well. Those relations are also $\approx 10\%$ higher than in our fiducial cosmology. At higher $z$, the simulation data seem to show a small $\approx 5\%$ residual excess over our model, again within the uncertainty of the model and our simulation results.

Finally, Figure \ref{fig:match_pl} shows a comparison of our model predictions and the \cnurs for the self-similar cosmologies. Due to the lower weight given to these models in our parameter fit, the agreement is somewhat worse than for the $\Lambda$CDM cosmologies, about $15\%$ for the steeper slopes. In particular, we note that the self-similar models prefer a steeper \cnur at low $\nu$, i.e. a larger value of $\alpha$. Such discrepancies could of course be fixed at the expense of more free parameters. However, given that self-similar models are only of academic interest, and that the exact values of concentration can vary by at least $\approx$ 10--15$\%$ due to fitting methods, binning, and other factors, the achieved accuracy is more than sufficient.

%--------------------------------------------------------------------------------------------
\subsection{Model Predictions for High-redshift Micro-halos}
\label{sec:discussion:highz}
%--------------------------------------------------------------------------------------------

\begin{figure}
\centering
\includegraphics[trim = 5mm 12mm 0mm 3mm, clip, scale=0.78]{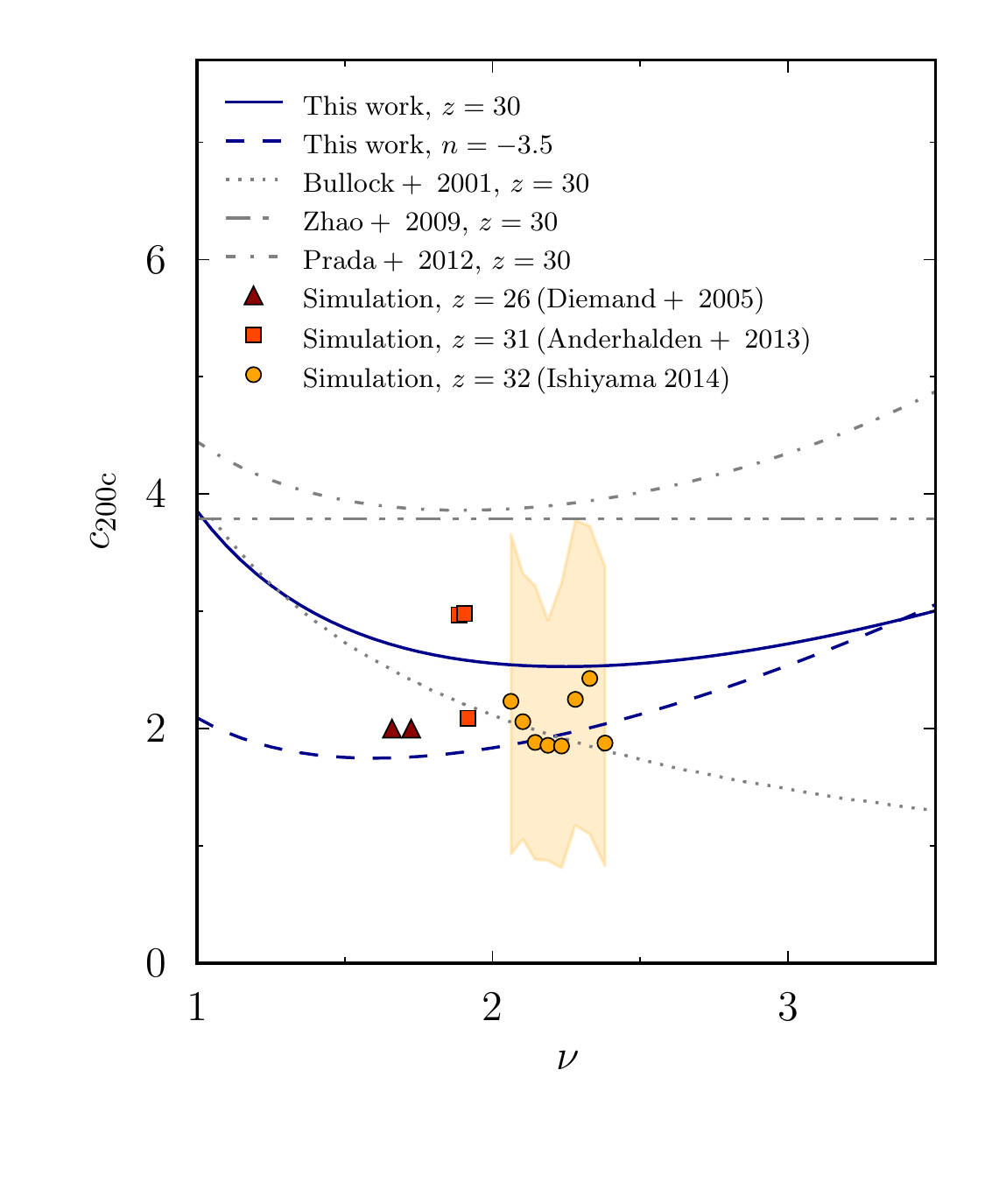}
\caption{Concentrations of micro-mass halos at high redshifts measured in cosmological simulations and predicted by different models. The solid blue line shows the prediction of our model at $z = 30$ for our fiducial $\Lambda$CDM cosmology. At such high redshift, the size scales involved are small, $k_{\rm R}$ is large, and the power spectrum approaches a constant slope of about $-2.9$ (Figure \ref{fig:pkslope}). For some of the simulations, the initial power spectrum had an explicit cutoff at the free-streaming scale and is thus not well described by the \citet{eisenstein_98} approximation. Near this cutoff, the effective slope could easily be as steep as $n = -3.5$ (the model prediction for this slope is shown in the dashed blue line). The simulation results from \citealt{diemand_05} refer to their quoted concentrations for two individual halos (triangles). The results of \citealt{anderhalden_13} (squares) correspond to their simulation without a power spectrum cutoff. The shaded area around the \citealt{ishiyama_14} results (circles) indicates the 33\% and 66\% range. The predictions of the \citet{bullock_01_halo_profiles}, \citet{zhao_09_mah}, and \citet{prada_12} models at $z = 30$ are plotted for comparison (gray lines). See Section \ref{sec:discussion:highz} for a detailed discussion.}
\label{fig:highz}
\end{figure}

The concentration model presented above was calibrated using simulation results for relatively massive halos expected to host galaxies. However, it is interesting to test whether the model correctly extrapolates to simulation results outside this regime. For example, the \cmr is often modeled with power-law functions that extrapolate to very high concentrations for the smallest, Earth--mass halos. In contrast, power-law functions in \cnu space lead to a flattening of the \cmr at low masses, thereby predicting much smaller concentrations in the low-mass regime \citep[see, e.g., Figure 10 in][]{ludlow_14_cm}. 

In Figure \ref{fig:highz} we compare the concentrations measured in simulations of micro-halos reported by various groups, as well as the predictions of our model. The halo masses range from $2 \times 10^{-7} \msunh$ to $10 \msunh$, up to $16$ orders of magnitude smaller than the smallest halos used to calibrate the model. In the three numerical studies we compare to, the density profiles are fit with an extended NFW profile with a variable inner slope. In \citet{anderhalden_13} and \citet{ishiyama_14}, a conversion to standard NFW concentration is provided. The profile fitted to the halos in \citet{diemand_05} has an inner slope of $\alpha = -1.2$. We convert the given concentration ($c = 1.6$) using the formula of \citet{ricotti_03}, $c_{\rm NFW} = c_\alpha / (2-\alpha)$, giving $c_{\rm NFW} = 2$. The results of the three different studies are consistent with each other and show that micro--mass halos have low concentrations, $1 \lsim c \lsim 3$, with no discernible dependence on halo mass.

The solid blue line in Figure \ref{fig:highz} shows the prediction of our model at $z = 30$ which matches the simulation results reasonably well. At the very high redshift and very small scales we are considering, the power spectrum slope approaches a nearly constant value of $-2.9$ in the case of the fiducial cosmology (Figure \ref{fig:pkslope}). Thus, the prediction becomes similar to predictions at fixed slope $n$, while the exact cosmology matters relatively little. \citet{diemand_05} state that the effective slope of their power spectrum near the cutoff is about $-3$. As some of the simulations plotted in Figure \ref{fig:highz} have a power spectrum with a cutoff corresponding to the free streaming scale of dark matter particles, the effective slope experienced by halos at the cutoff may be even steeper than $-3$. For illustration, the dashed line in Figure \ref{fig:highz} shows the predictions of our model for a fixed slope of $-3.5$ which leads to even lower concentrations than our $z = 30$ prediction. Regardless of the exact slope, our model predicts low concentrations, $c < 3$, across a wide range of masses, and a shallow, rising \cnur, in excellent agreement with the simulation results. We have thus validated our model across $22$ orders of magnitude in mass, and from $z = 0$ to $z \approx 30$. We discuss the other models shown in Figure \ref{fig:highz} in Section \ref{sec:discussion:comp2}.

\begin{figure*}[!ht]
\centering
\includegraphics[trim = 6mm 18mm 2mm 4mm, clip, scale=0.7]{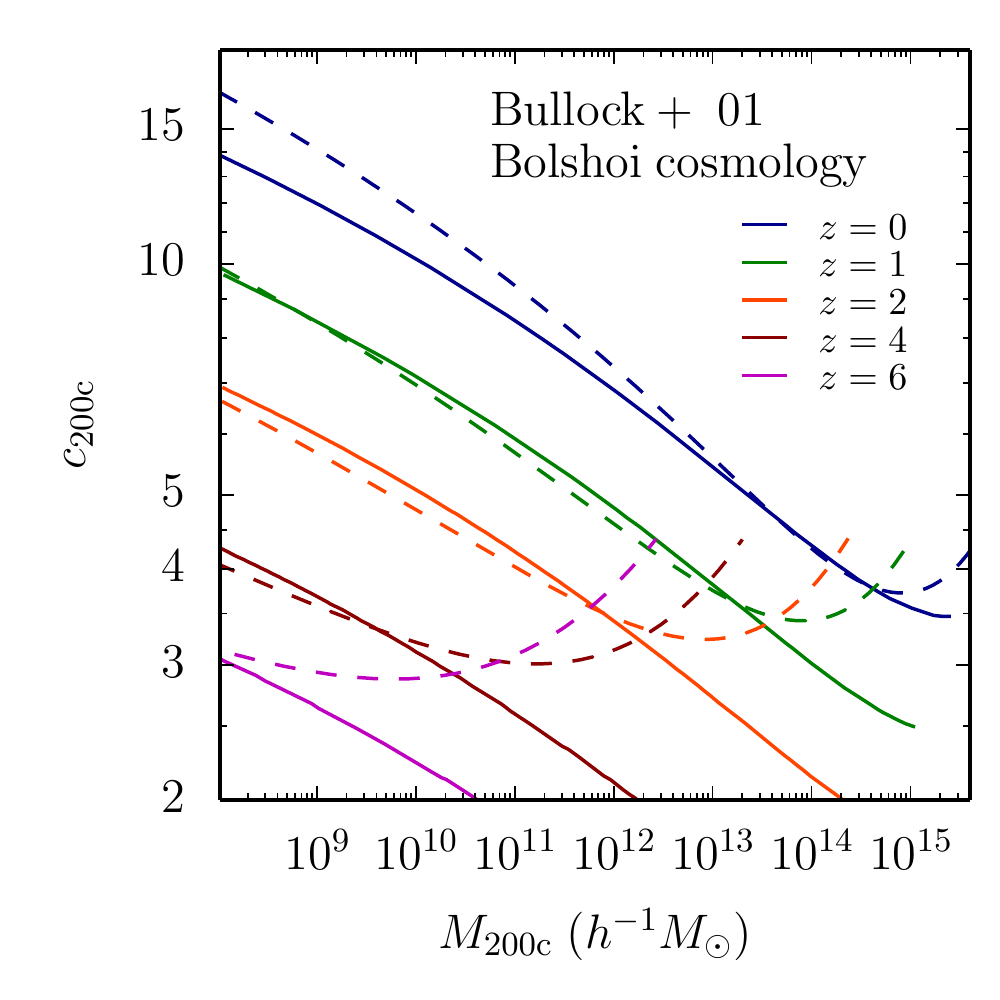}
\includegraphics[trim = 20mm 18mm 2mm 4mm, clip, scale=0.7]{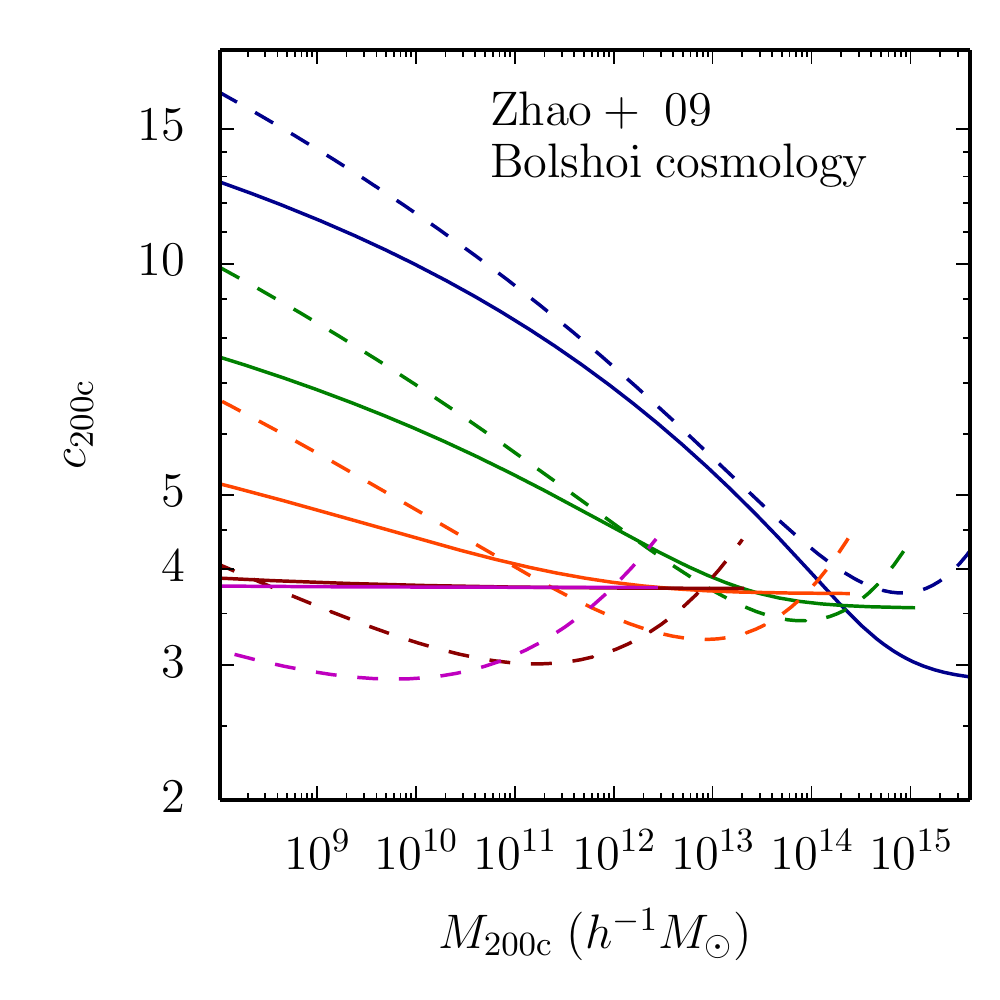}
\includegraphics[trim = 20mm 18mm 2mm 4mm, clip, scale=0.7]{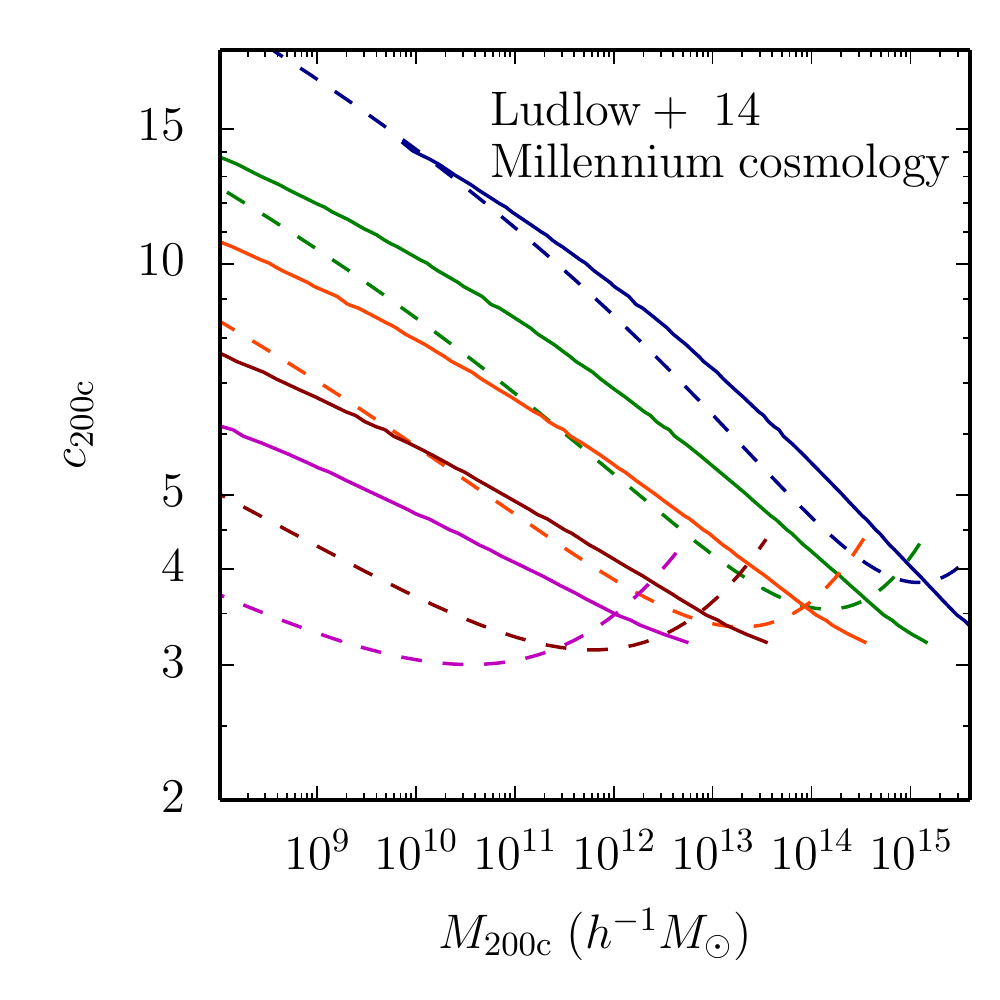}
\includegraphics[trim = 6mm 3mm 2mm 4mm, clip, scale=0.7]{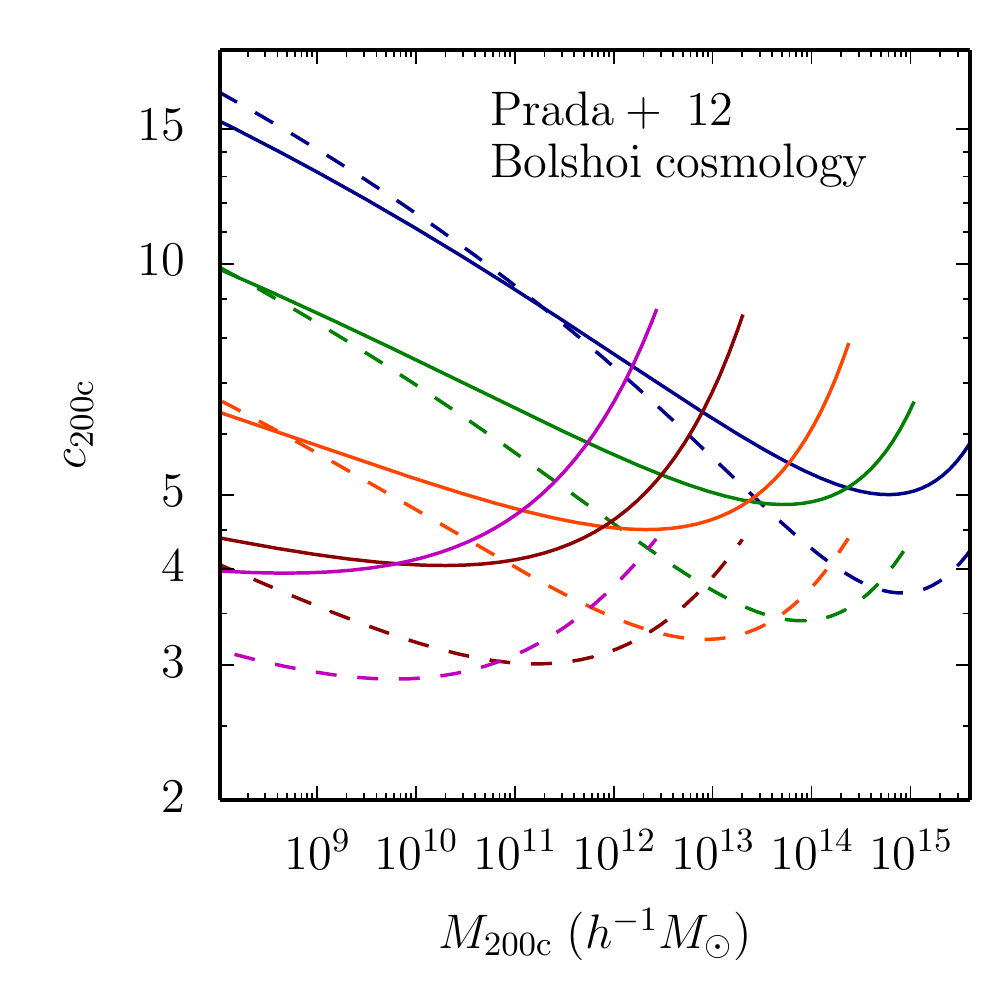}
\includegraphics[trim = 20mm 3mm 2mm 4mm, clip, scale=0.7]{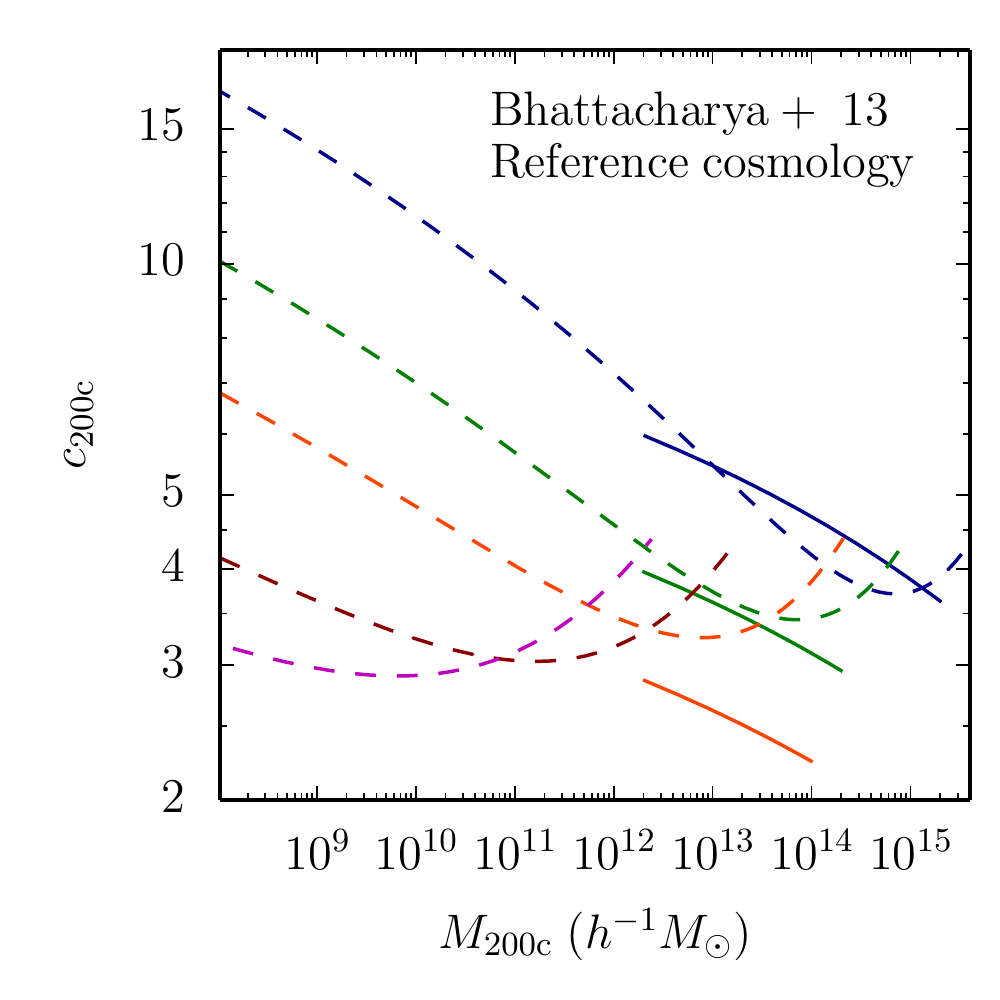}
\includegraphics[trim = 20mm 3mm 2mm 4mm, clip, scale=0.7]{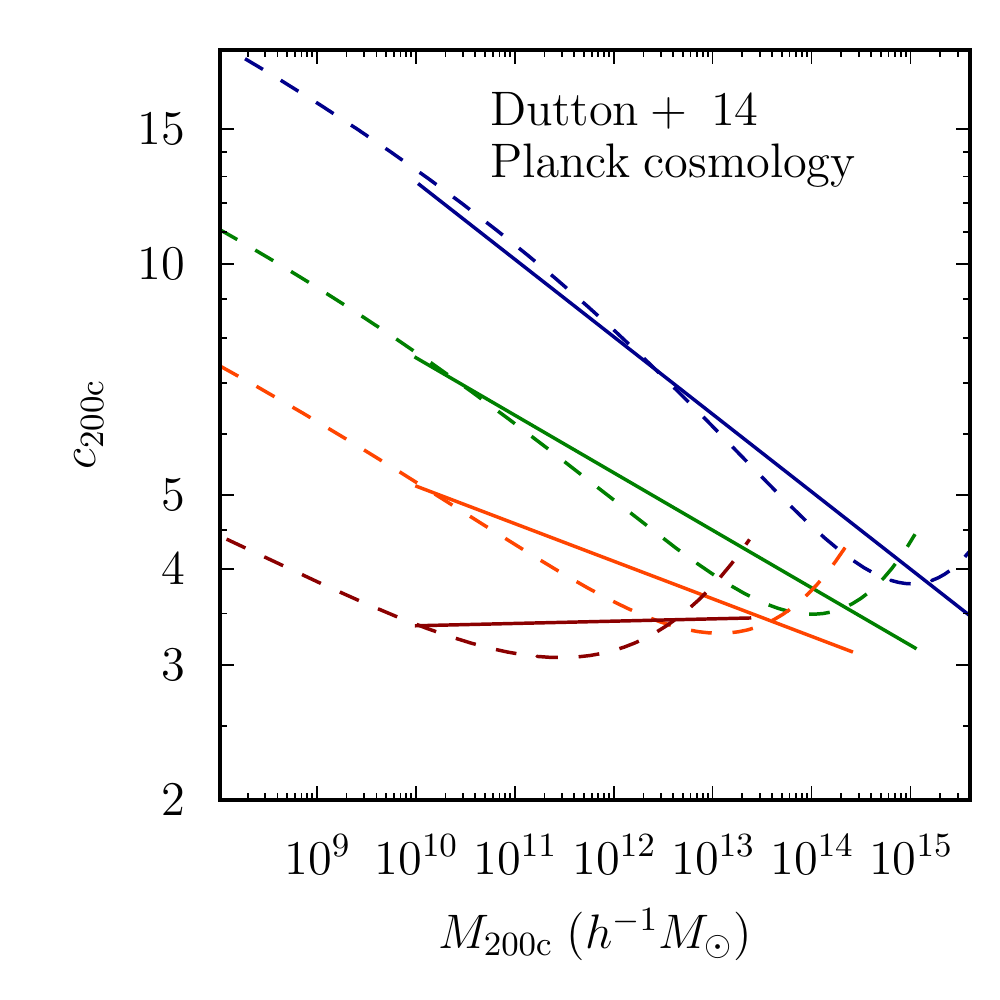}
\caption{Comparison of our model (dashed lines) with models from the literature (solid lines), namely the models of \citet{bullock_01_halo_profiles}, \citet{zhao_09_mah} and \citet{ludlow_14_cm} in which concentration is based on the MAH of halos, the \cnu models of \citet{prada_12} and \citet{bhattacharya_13}, and the power-law fitting function of \citet{dutton_14}. Some of these models were calibrated for cosmologies different from our fiducial cosmology. In those cases, the dashed lines show the predictions of our model for the respective cosmology. The \citet{bhattacharya_13} model is only plotted in the mass range where it was calibrated. See text for a detailed discussion of the differences between the models.}
\label{fig:modelcomparison}
\end{figure*}

%--------------------------------------------------------------------------------------------
\section{Discussion}
\label{sec:discussion}
%--------------------------------------------------------------------------------------------

We have presented a universal model for halo concentrations in which concentration is a function of two variables: peak height, and the local slope of the matter power spectrum. Both dependencies are motivated by physical arguments. Our results demonstrate that these two variables and seven fitted parameters are sufficient to explain the behavior of halo concentrations across a wide range of masses, redshifts, and cosmologies, including scale--free $\Omega_{\rm m} = 1$ cosmologies. In this section we compare our results with those of previous numerical studies (Section \ref{sec:discussion:comp1}) and general concentration models previously proposed in the literature (Section \ref{sec:discussion:comp2}). Finally, we compare our results with current observational estimates of halo concentrations on cluster mass scales (Section \ref{sec:discussion:observations}).

%--------------------------------------------------------------------------------------------
\subsection{Comparison with Previous Numerical Calibrations}
\label{sec:discussion:comp1}
%--------------------------------------------------------------------------------------------

Our simulation results generally agree with the recent study of \citet[][compare, for example, their Figure 14 and our Figure \ref{fig:match_bolshoi}]{dutton_14}.
They conclude that concentrations in the Planck cosmology are $\approx 20\%$ larger than in the $WMAP5$ cosmology, which is consistent with our findings. The comparison in the bottom right panel of Figure \ref{fig:modelcomparison} shows a good overall agreement of \citet{dutton_14} and the predictions of our model for the same Planck cosmology. Considering that the results were obtained using different $N$-body codes, halo finders, and were subject to different resolution limits, the good agreement is reassuring.

However, there are also important differences: at higher $z$, the power-law fits become progressively poorer descriptions of the shape of the \cnur over the mass range probed. Our model approaches a power-law in \cnu space at low $\nu$, and has an upturn at high $\nu$. In contrast, a power-law in \cm space extrapolates to large concentrations at small masses, and small concentrations at large masses. The high-$z$ results in Section \ref{sec:discussion:highz} highlight the low-mass issue in particular. Many other studies have used power-law fits to the \cmrs measured in simulations as a compact way to approximate their numerical results \citep{jing_00_profiles2, dolag_04, neto_07, duffy_08, gao_08, maccio_08, klypin_11_bolshoi, munozcuartas_11}. These calibrations will likewise be inaccurate at low and high masses. 

For this reason, both \citet{prada_12} and \citet{ludlow_14_cm} have recently advocated modeling the \cnur rather than the \cmr, and \citet{bhattacharya_13} have approximated the concentrations in their simulations using power-law fits in \cnu space (bottom center panel of Figure \ref{fig:modelcomparison}). Their fits were calibrated for a  wide range of cosmologies, but only for high masses, $M > 2 \times 10^{12} \msunh$, and clearly cannot be extrapolated to low masses because the \cnur significantly steepens at low $\nu$. Overall, our results clearly show that a power-law is not a good approximation to the \cnur over a wide range of $\nu$. Furthermore, \citet{bhattacharya_13} find much stronger deviations of the \cnur from universality than we do. For example, they observe a $\approx 30\%$ difference in normalization between $z = 0$ and $z = 2$ for massive halos (their Figure 2). This difference leads to a strong redshift evolution of their fitting function, in disagreement with both our model and the results of \citet{dutton_14}. The reason for this discrepancy is not entirely clear, but we note that it arises primarily at $z > 1$ (Figure \ref{fig:modelcomparison}).

\begin{figure*}
\centering
\includegraphics[trim = 0mm 0mm 2mm 2mm, clip, scale=0.66]{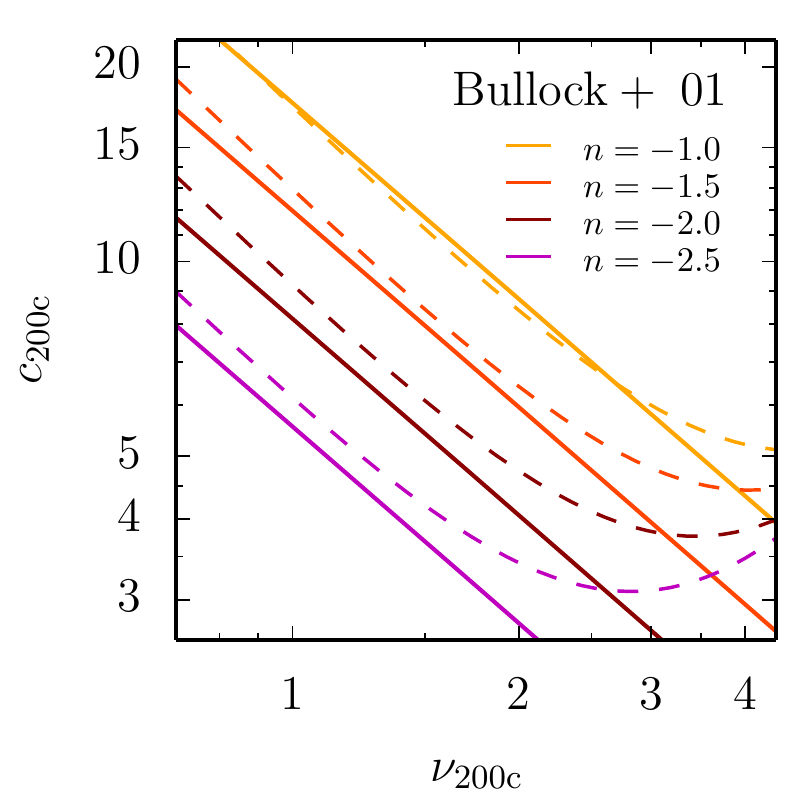}
\includegraphics[trim = 16mm 0mm 2mm 2mm, clip, scale=0.66]{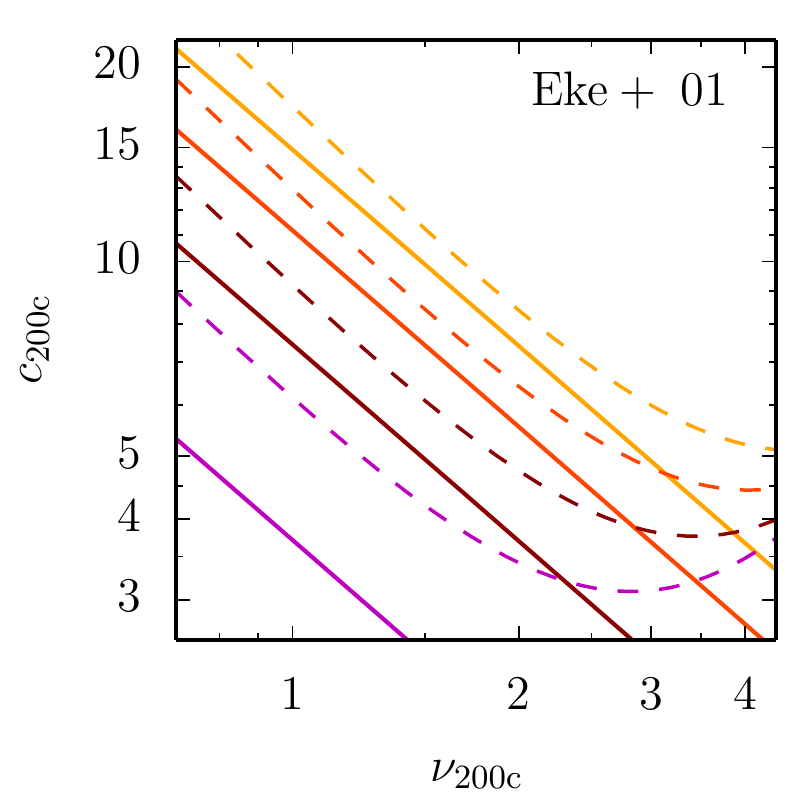}
\includegraphics[trim = 16mm 0mm 2mm 2mm, clip, scale=0.66]{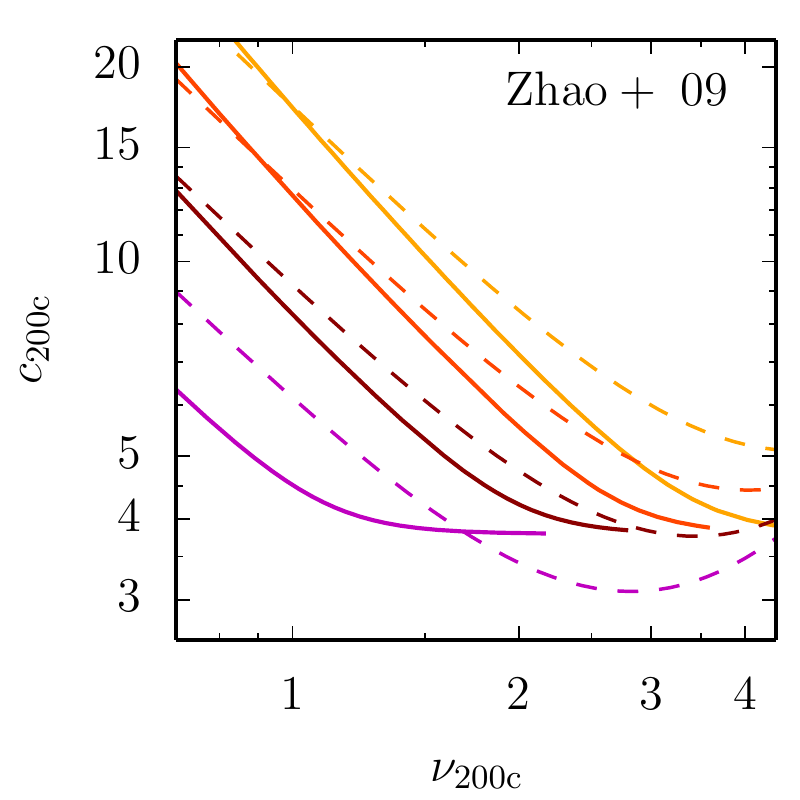}
\includegraphics[trim = 16mm 0mm 2mm 2mm, clip, scale=0.66]{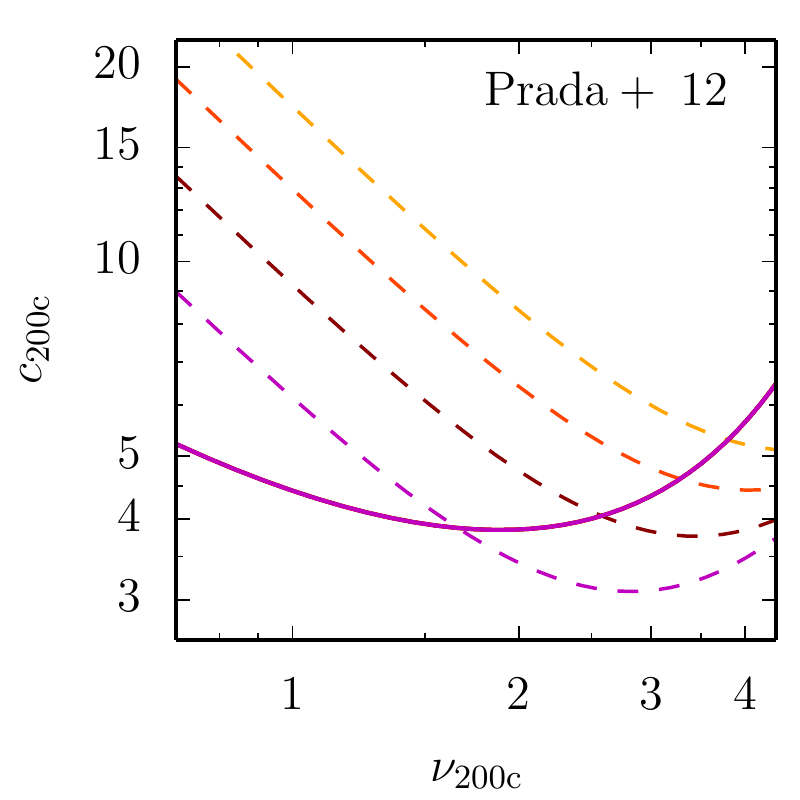}
\caption{Comparison of our model for self-similar cosmologies (dashed lines) with models from the literature (solid lines), namely the models of \citet{bullock_01_halo_profiles}, \citet{eke_01_concentrations}, \citet{zhao_09_mah}, and \citet{prada_12}. Both the \citet{bullock_01_halo_profiles} and \citet{eke_01_concentrations} models predict $\cvir \propto 1 / \nuvir$, but a different dependence on $n$. The \citet{prada_12} model predicts no dependence on $n$. See text for a detailed discussion.}
\label{fig:modelcomparison_pl}
\end{figure*}

We measure the 68\% rms scatter in $\log_{10}c_{200c}$ around the median concentration and find it to be $\approx 0.16$ dex, independent of redshift, peak height, or mass definition. This value is in excellent agreement with various previous measurements \citep[e.g.,][]{bullock_01_halo_profiles, wechsler_02_halo_assembly, duffy_08, bhattacharya_13}. Some authors have measured lower values of $\approx 0.10$ dex \citep{maccio_08, dutton_14}, but for samples that included only relaxed objects \citep{bhattacharya_13}. We note that we have not corrected the scatter for any errors in the concentration measurements, i.e., uncertainties in the best-fit parameters of the NFW profile. The measured scatter is thus an upper limit on the true scatter in concentration.

%--------------------------------------------------------------------------------------------
\subsection{Comparison with Previous Concentration Models}
\label{sec:discussion:comp2}
%--------------------------------------------------------------------------------------------

While the power-law fits discussed in the previous section allow a simple and compact parameterization of simulation results, they extrapolate inaccurately outside the range of redshifts, masses, and cosmologies over which they were calibrated. A number of authors have thus proposed more physically motivated models of concentration, calibrated using simulations. Most of these models have used the tight coupling between concentrations and halo MAHs \citep{navarro_97_nfw, bullock_01_halo_profiles, wechsler_02_halo_assembly, zhao_03_mah, giocoli_12}. Here we compare our results in detail to four such models \citep{bullock_01_halo_profiles, eke_01_concentrations, zhao_09_mah, ludlow_14_cm}, and to the empirical \cnu model of \citet{prada_12}. We consider both $\Lambda$CDM (Figure \ref{fig:modelcomparison}) and self-similar cosmologies (Figure \ref{fig:modelcomparison_pl}). While the latter models do not represent the real universe, they provide an interesting test of the universality of any model that relies on $P(k)$ for its predictions, and thus any model that works with $\sigma(R)$, $M^*$, or $\nu$.

The model of \citet[][we use the improvements to this model proposed by \citealt{maccio_08}]{bullock_01_halo_profiles} predicts a simple scaling with redshift, $\cvir \propto a/a_{\rm c}$, where $a_{\rm c}$ is the expansion factor at the epoch when the halo assembled a certain fraction of its mass. The comparison with our model shows that this model captures the \cmr at $z = 0$ well, and also predicts the correct redshift evolution at low $\nu$ where it was calibrated. However, the model does not reproduce the upturn or even a flattening at high $\nu$ and $z > 0$, which is clearly visible in our results and other recent studies. Similarly, for self-similar cosmologies the \citet{bullock_01_halo_profiles} model matches our simulation results at low $\nu$ but fails at high $\nu$. Thus, although Figure \ref{fig:highz} shows that the \citet{bullock_01_halo_profiles} model predicts the concentrations of  micro-mass halos at $z = 30$ correctly, this agreement may be coincidental, as the model does not match halo concentrations at similar peak heights at $z\lesssim 6$. For example, at $\nu \gsim 2$ and $z = 6$ (corresponding to $M > 10^9\ M_{\odot}$), the prediction of the \citet{bullock_01_halo_profiles} model does not match our results  (see Figure \ref{fig:modelcomparison}). 

\citet{eke_01_concentrations} expanded on the models of \citet{navarro_97_nfw} and \citet{bullock_01_halo_profiles} by adding an explicit dependence on the power spectrum slope via a term proportional to $d \log \sigma(M) / d \log M$, and a similar dependence is already implicit in the \citet{bullock_01_halo_profiles} model. The two main differences between these models and ours are that instead of the slope of $\sigma(M)$ we consider the slope of $P(k)$, and that in their models the power spectrum slope influences concentration through the formation redshift of a halo. As a result, $n$ significantly changes the normalization of the \cmr, but its shape varies relatively little with $n$. In contrast, in our model both the normalization and shape of the relation explicitly depend on $n$ (Figure \ref{fig:modelcomparison}). These differences are particularly apparent in the predictions for self-similar cosmologies (Figure \ref{fig:modelcomparison_pl}). Both the \citet{bullock_01_halo_profiles} and \citet{eke_01_concentrations} models predict a power-law shape, while the actual \cnur at high $\nu$ flattens and even turns up. In addition, the model of \citet{eke_01_concentrations} predicts a stronger dependence of the normalization on $n$ than we observe in our scale-free simulations. 

In the model of \citet{zhao_09_mah}, concentration is a function of the time since a halo accumulated $4\%$ of its mass, with a floor of $\cvir \geq 4$ (corresponding to $\ctoc \gsim 3.8$). Thus, all halos in the fast accretion regime (i.e., during their early evolution) have concentrations around $4$, whereas their concentration increases later in the slow accretion regime. While our models agree fairly well at low $z$, they diverge at higher $z$ where the \citet{zhao_09_mah} model predicts a mass-independent $\cvir \approx 4$. Our results \citep[see also][]{dutton_14} show that there is no well-defined floor in the concentration values: halos that form from the collapse of perturbations with a steep power spectrum have concentrations much smaller than the floor value adopted by \citet{zhao_09_mah}. A similar picture emerges for the self-similar cosmologies, where the model predicts virtually no $n$-dependence at high $\nu$, and is thus incompatible with our simulation results. 

\citet{ludlow_14_cm} have recently argued that concentration does not only reflect the formation epoch of a halo \citep[as previously advocated by][]{wechsler_02_halo_assembly, zhao_03_mah}, but that the entire density profile of a halo is a one-to-one reflection of the critical density of the universe when different parts of the halo were assembled. Based on this insight, as well as an analytic prescription for the assembly history of halos, they propose a \cnu model that agrees with ours relatively well at $z = 0$. At high redshifts, however, their model does not match our simulation results because they assume the \cnur to be universal in redshift. Their model also does not exhibit the upturn at high masses that is apparent in our simulations \citep[see also][]{prada_12, dutton_14}, presumably because they only consider relaxed halos.

Finally, we compare our model to that of \citet{prada_12} which, unlike the models discussed so far, is not based on the formation time of halos. Instead, they propose a fitting formula with $13$ free parameters to the \cnur in their $\Lambda$CDM simulations and its redshift dependence. The bottom left panel of Figure \ref{fig:modelcomparison} shows large differences between our models, which probably arise because we estimate concentrations from direct fits to the mass profile while \citet{prada_12} derive them from the maximum circular velocity, $v_{\rm max}$, of halos and assume the NFW form. They also bin concentrations in $v_{\rm max}$ rather than in mass. Both choices have been shown to significantly affect the resulting \cmr \citep{meneghetti_13}. In addition to the large differences in normalization and slope, the upturn at high $\nu$ and high $z$ is weaker in our simulations. However, the main difference between our models is in the physical mechanism invoked to explain the non-universality of the \cnur. While \citet{prada_12} make an empirical redshift correction based on the linear growth factor, our model explains the non-universality using $n$ as a second parameter in addition to $\nu$. For the self-similar cosmologies (Figure \ref{fig:modelcomparison_pl}), the \citet{prada_12} model predicts no dependence on $n$, and thus fails to reproduce our simulation results. 

\citet{sanchezconde_14} recently investigated the predictions of the \citet{prada_12} model for micro-halos, and concluded that the model is in good agreement with simulation results in \cm space, in contradiction with the comparison shown in our Figure \ref{fig:highz}. However, their conclusion appears to be based on the top panel of their Figure 1 in which high-$z$ results for $c_{200c}$ are rescaled to $z = 0$ using the $c \propto a$ scaling of \citet{bullock_01_halo_profiles}. This scaling was derived for $c_{\rm vir}$ and is inaccurate for $c_{200c}$ (see Appendix \ref{sec:discussion:evolution}). When this correction is not applied (bottom panel of their Figure 1), the model of \citet{prada_12} does indeed predict concentrations a factor of two higher than the simulation results, consistent with our findings.

In conclusion, none of the universal models previously proposed in the literature can explain our simulation results over the full range of masses and redshifts probed. Particularly, at high masses and high redshifts the assumptions of many models are too simplistic, e.g., a concentration floor at high masses, or no flattening of the relation at all. Recently, there has been significant debate regarding the high-$\nu$ behavior of the \cnur \citep{prada_12, ludlow_12, bhattacharya_13, meneghetti_13, ludlow_14_cm}. In particular, the exponential upturn detected by \citet{prada_12} was traced to their binning scheme and the $v_{\rm max}$ approximation \citep{meneghetti_13}. The largest halos are often the least well fit by NFW profiles \citep{duffy_08, diemer_14_profiles}, leading to the greatest differences between scale radii estimated using the $v_{\rm max}$ approximation and those derived from a profile fit. Furthermore, \citet{ludlow_12} showed that the upturn is due to unrelaxed halos. These differences seemed to explain why the model of \citet{prada_12} differs from all other models, particularly at the high-$\nu$ end. 

Nevertheless, we also find strong evidence for an upturn at high $\nu$, i.e. fits of Equation (\ref{eq:fitfunc}) with $\beta \leq 0$ result in a very poor match with our simulated concentrations. While there is no clear evidence for an upturn in the low-$z$ data for $\Lambda$CDM cosmologies (Figure \ref{fig:cm_mdefs}), the high-$z$ relations clearly take on a positive slope. The same is true for the low-$n$ self-similar cosmologies. \citet{ludlow_12} showed that the upturn can be removed, but only if a large fraction of all halos is excluded as unrelaxed, particularly at high $z$ (see their Table 1). Since we choose to consider the full halo sample, our model does predict an upturn. This feature is particularly salient in regimes where halos form in regions with a steep slope of the power spectrum, which leads to halos of different masses forming at a similar time, and thus a higher fraction of unrelaxed halos. While the NFW profile does not fit those objects as well as slowly accreting objects \citep{diemer_14_profiles}, the 68\% scatter in concentration does not increase at high mass, indicating that the concentrations in this regime are measured with similar accuracy as for relaxed objects.

%--------------------------------------------------------------------------------------------
\subsection{Comparison with Observations}
\label{sec:discussion:observations}
%--------------------------------------------------------------------------------------------

\begin{figure}
\centering
\includegraphics[trim = 2mm 3mm 5mm 2mm, clip, scale=0.75]{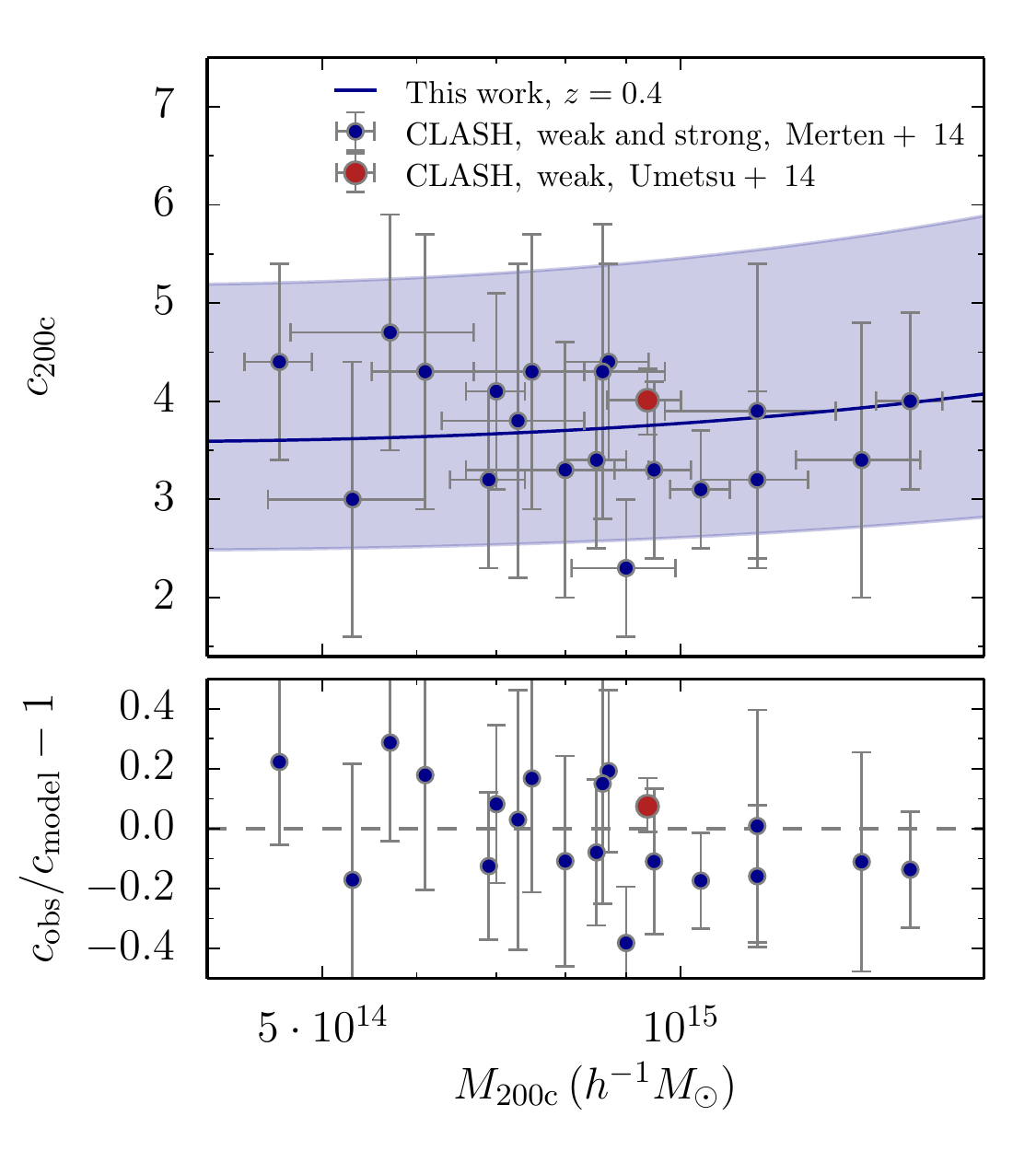}
\caption{Comparison of our model predictions with the concentration measurements for clusters in the CLASH sample, derived using the weak lensing measurement of \citet[][for all CLASH clusters, red point]{umetsu_14} and the strong and weak lensing measurements of \citet[][for individual clusters, blue points]{merten_14}. Our model was evaluated at the mean redshift of the CLASH clusters, $z = 0.4$, and for the cosmology assumed in the CLASH papers (blue line, the shaded are indicates a 68\% scatter of 0.16 dex). The bottom panel shows the residuals between the measured concentrations and our model. For this comparison, our model was evaluated at the redshift of each cluster, and at $z = 0.35$ for the weak lensing measurement.}
\label{fig:observations}
\end{figure}

Measuring halo concentrations in observed systems is challenging as it requires accurate measurements of the dark matter density profile. The most accurate measurements have been derived from either X-ray or lensing observations of clusters of galaxies, but the results initially appeared to be in tension with simulations. For example, the X-ray results of \citet{schmidt_07} and \citet{ettori_10} seemed to indicate a much steeper slope of the \cmr than measured in simulations. However, \citet{rasia_13} pointed out that there are many factors that can lead to such disagreement, for example baryonic effects, deviations from the assumption of hydrostatic equilibrium, or the X-ray selection function. Similarly, lensing observations seemed to indicate an ``overconcentration'' problem, with low-mass clusters having much higher concentrations than predicted by simulations, and thus a steeper overall \cmr than expected \citep[e.g.,][]{oguri_12, wiesner_12}. Recently, however, \citet{auger_13} pointed out that the observed steepness of the \cmr is an artifact of neglecting the covariance between the errors in mass and concentration. 

A good agreement between observations and simulations was recently reported for the results of the CLASH cluster lensing survey \citep{umetsu_14, merten_14}. When the X-ray selection function is taken into account, the \cmr measured in their simulations \citep{meneghetti_14} matches the observations very well. The CLASH simulations were non-radiative and thus do not reproduce important baryonic effects which can change the total mass profile of halos \citep[e.g.,][]{rudd_08}. Furthermore, the translation between $N$-body and hydrodynamic simulations is more complicated than a simple shift in the \cmr \citep{velliscig_14}. Nevertheless, the good agreement between the CLASH observations and simulations indicates that we can attempt at least a rough comparison between the CLASH results and our model, even though our model (as well as the models from the literature discussed in the previous sections) are based on pure dark matter simulations. In any case, the uncertainties in the observational data are still significantly larger than baryonic effects \citep{meneghetti_14}.  

Figure \ref{fig:observations} shows our model (blue line), evaluated at the mean redshift of the CLASH clusters, $z = 0.4$. In their weak-lensing analysis of $16$ X-ray selected CLASH clusters, \citet{umetsu_14} find a mean concentration of $\ctoc = 4.01^{+0.35}_{-0.32}$ at an effective halo mass of $9.38^{+0.70}_{-0.63} \times 10^{14} \msunh$ and a mean redshift of $z = 0.35$ (red point), in good agreement with our model which predicts $\ctoc = 3.73$, within $1 \sigma$ of the measured result. \citet{merten_14} analyze both the weak and strong lensing signals of 18 CLASH clusters, and find masses and concentrations for each individual cluster (blue points). For the comparison in the bottom panel of Figure \ref{fig:observations}, our model was evaluated at the redshift of each individual cluster. The \citet{merten_14} concentrations are distributed around $\ctoc = 3.7$ with a weak mass dependence, in excellent agreement with our model. 

In conclusion, within the current accuracy of lensing observations, there is no evidence for strong, $\gtrsim 20\%$, baryonic effects on the concentrations of clusters. Note, however, that the scatter of the individual cluster concentrations around the mean is about $0.08$ dex, significantly smaller than what we measure for simulated halos. However, as we noted previously, our scatter estimate includes errors in the concentration measurement and is thus an upper limit of the true scatter. Detailed comparisons of the predicted and observed scatter will demand larger, mass-selected samples and more careful estimates of the fit errors in simulation analyses.

%--------------------------------------------------------------------------------------------
\section{Conclusions}
\label{sec:conclusion}
%--------------------------------------------------------------------------------------------

An accurate calibration of halo concentrations as a function of mass, redshift, and cosmology is important for our understanding of halo structure and interpretation of observations. We have presented a numerical study of halo concentrations in $\Lambda$CDM and self-similar cosmologies, and a universal model that accurately describes our simulation results across the entire range of masses, redshifts, and cosmologies we explored, including scale-free $\Omega_{\rm} = 1$ cosmologies, as well as the concentrations of Earth--mass halos. Our main conclusions are as follows.
\begin{enumerate}
\item The relation between concentration and peak height exhibits the smallest deviations from universality for halo radii defined with respect to the critical density of the universe. Definitions using the virial density contrast, or a contrast relative to the mean density, result in much larger deviations from universality. 

\item Our simulations show that both the normalization and shape of the \cnur depend on the local slope of the matter power spectrum, $n$. In particular, we find that there is no well-defined floor in the concentration values. Instead, the minimum concentration value depends on redshift: at fixed $\nu$, a higher $z$ corresponds to steeper values of $n$, and lower minimum concentrations. The \cnur for steep spectral slopes exhibits a well-defined upturn at high $\nu$, which is likely associated with an increased fraction of unrelaxed halos in such a regime. 

\item We show that concentrations can be described as a function of only two parameters, peak height, $\nu$, and the slope of the linear matter power spectrum, $n$. In $\Lambda$CDM cosmologies, we define $n$ as the local slope of the power spectrum at a scale close to the Lagrangian radius of a halo.

\item We present a seven-parameter, double power-law functional form approximating the $c(\nu, n)$ relation which can easily be evaluated for any known power spectrum. This function fits concentrations in the fiducial $\Lambda$CDM cosmology to $\lesssim 5\%$ accuracy, and those in scale-free $\Omega_{\rm m} = 1$ models to $\lesssim 15\%$ accuracy. The model predicts the low concentration values of Earth--mass micro-halos at $z \approx 30$, and thus correctly extrapolates over $16$ orders of magnitude in halo mass. 

\item The predictions of our model significantly differ from all models previously proposed in the literature at high masses and redshifts. For lower masses, we find that our results are well approximated by the model of \citet{bullock_01_halo_profiles}.

\item The predictions of our model for the average concentrations of cluster halos are in excellent agreement with the recent observational measurements from the CLASH cluster lensing survey. 
\end{enumerate}
We provide a public, stand-alone Python code to evaluate the predictions of our model for arbitrary cosmologies at \url{benediktdiemer.com/code}. We also provide tables of concentration as a function of mass and redshift for a large range of cosmologies at the same website under \url{benediktdiemer.com/data}.

%%%%%%%%%%%%%%%%%%%%%%%%%%%%%%%%%%%%%%%%%%%%%%%%%%%%%%%%%%%%%%%%%%%%%%%%%%
% ACKNOWLEDGMENTS
%%%%%%%%%%%%%%%%%%%%%%%%%%%%%%%%%%%%%%%%%%%%%%%%%%%%%%%%%%%%%%%%%%%%%%%%%%

\vspace{0.5cm}

We are grateful to Matt Becker for his assistance with running the simulations, and for making his \textsc{CosmoPower} code public. Likewise, we thank Peter Behroozi for making his \textsc{Rockstar} halo finder code publicly available, and Andrew Hearin and Sam Skillman for testing our public code. We thank Neal Dalal, Surhud More and Keiichi Umetsu for many useful discussions. This work was supported by NASA ATP grant NNH12ZDA001N and by the Kavli Institute for Cosmological Physics at the University of Chicago through grants NSF PHY-0551142 and PHY-1125897 and an endowment from the Kavli Foundation and its founder Fred Kavli. We have made extensive use of the NASA Astrophysics Data System and {\tt arXiv.org} preprint server. Most of the simulations used in this study have been carried out using the {\tt midway} computing cluster supported by the University of Chicago Research Computing Center.

%%%%%%%%%%%%%%%%%%%%%%%%%%%%%%%%%%%%%%%%%%%%%%%%%%%%%%%%%%%%%%%%%%%%%%%%%%
% APPENDIX
%%%%%%%%%%%%%%%%%%%%%%%%%%%%%%%%%%%%%%%%%%%%%%%%%%%%%%%%%%%%%%%%%%%%%%%%%%

\appendix

%--------------------------------------------------------------------------------------------
\section{On the Redshift Evolution of Individual Halos}
\label{sec:discussion:evolution}
%--------------------------------------------------------------------------------------------

In this Appendix we briefly comment on the evolution of the concentrations of {\it individual} halos. \citet[][see also \citealt{wechsler_02_halo_assembly}]{bullock_01_halo_profiles} showed that halo concentrations evolve as $\cvir \propto a$ if they are defined using the radius enclosing a variable, ``virial'' overdensity. However, this scaling is sometimes used in the literature to describe other definitions of concentration. For example, a number of authors have used the scaling to translate very high redshift concentrations to $z = 0$ \citep{anderhalden_13, sanchezconde_14, ishiyama_14}. 

\begin{figure}
\centering
\includegraphics[trim = 5mm 0mm 0mm 0mm, clip, scale=0.75]{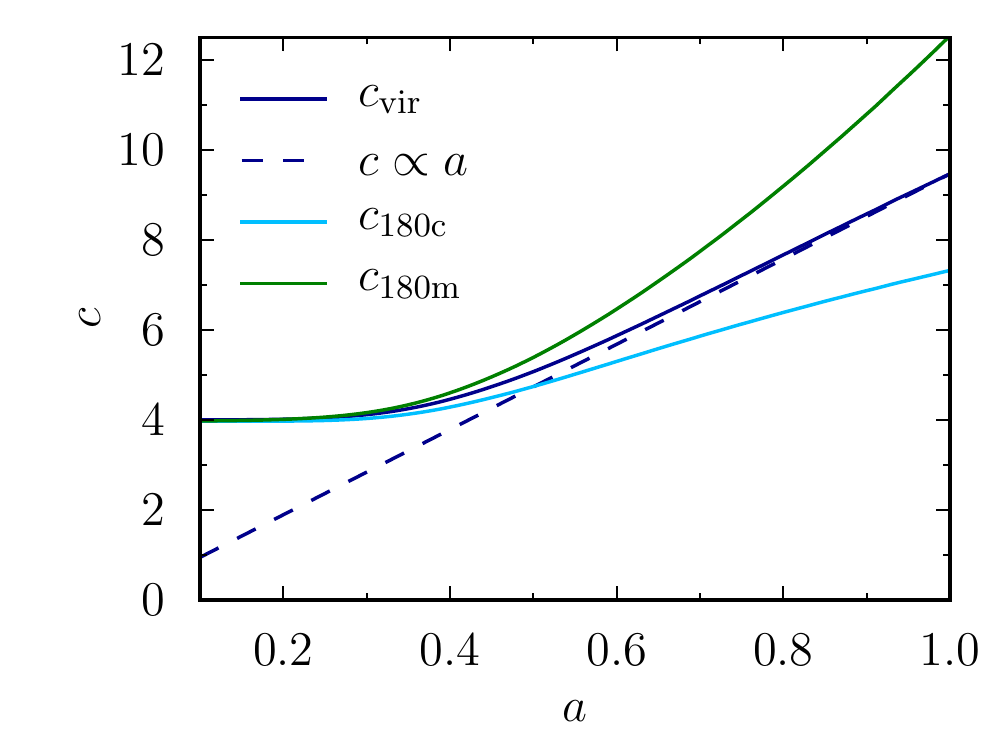}
\caption{Evolution of different definitions of concentration for the same physical halo. The dark blue line shows the evolution of $\cvir$ for a halo of mass $\mvir = 10^{12} \msunh$ at $z = 0$, as predicted by the \citet{zhao_09_mah} model. For the same physical halo, the green and light blue lines show the evolution of $c_{\rm 180m}$ and $c_{\rm 180c}$. While all three definitions of concentration share the same value at high redshift where the critical and mean densities of the universe are almost the same, they diverge significantly at low $z$. The frequently used scaling proposed by \citet{bullock_01_halo_profiles}, $c \propto a$ (dashed line), was designed specifically for $\cvir$ and is not a good fit to the evolution of the other concentration definitions. Furthermore, the scaling only works after the formation redshift, and can thus not be extrapolated to arbitrarily high $z$.}
\label{fig:evomdefs}
\end{figure}

Figure \ref{fig:evomdefs} demonstrates why the scaling is inappropriate for such applications. The figure shows the evolution of $\cvir$ (solid dark blue line) for a hypothetical halo with $\mvir = 10^{12} \msunh$ at $z = 0$, as predicted by the MAH model of \citet{zhao_09_mah}. At each redshift, we also compute $c_{\rm 180m}$ and $c_{\rm 180c}$ corresponding to the same physical density profile (green and cyan lines). The evolutions of the different concentration definitions are quite similar at high $z$, but diverge at $z\lesssim 2$, where $\Omega_{\rm m}<1$, and thus $\rhoc$ and $\rhom$ evolve differently. The dashed blue line shows the $\cvir \propto a$ scaling which is a good description of the evolution of $\cvir$, but {\it only for $\cvir$ and only after the formation redshift}. Thus, the scaling could only be applied to the evolution of $z = 30$ micro-halos if they stopped growing and merging at that redshift, which seems unlikely.

In \citet{diemer_13_pe}, we demonstrated that the evolution of the \cmr at low redshifts can almost entirely be explained by the pseudo-evolution of the outer halo radius, related to the evolution of the reference density, as opposed to real physical growth. Once halos enter the slow-accretion regime, their physical density profile, and thus their scale radius, barely change. Due to the decreasing reference density (critical or mean density of the universe), the virial radius, and thus concentration, grow. We checked that a simple model of a fixed, pseudo-evolving halo density profile describes the low-redshift evolutions shown in Figure \ref{fig:evomdefs} quite well. This agreement leads to the conclusion that the $\cvir \propto a$ scaling just {\it happens} to reproduce the pseudo-evolution of $\rvir$. 

Finally, we note a subtle secondary effect due to a combination of pseudo-evolution and cosmology. The critical density, in physical units, depends on $H_0$ and $\Omega_{\rm m}$, and hence differs between our cosmologies. Thus, for the same physical object, we measure a different $\rtoc$ and $\ctoc$. For example, the physical overdensity that corresponds to $200 \rhoc$ at $z = 0$ in the Bolshoi cosmology corresponds to $218 \rhoc$ in the Planck cosmology. We have quantified this effect and found it to change the \cnurs by a few percent. The redshift dependence is fairly complicated though, and we have chosen to ignore the issue.

%--------------------------------------------------------------------------------------------
\section{Peak Curvature as a Secondary Parameter}
\label{sec:discussion:curvature}
%--------------------------------------------------------------------------------------------

\begin{figure}
\centering
\includegraphics[trim = 2mm 2mm 5mm 0mm, clip, scale=0.73]{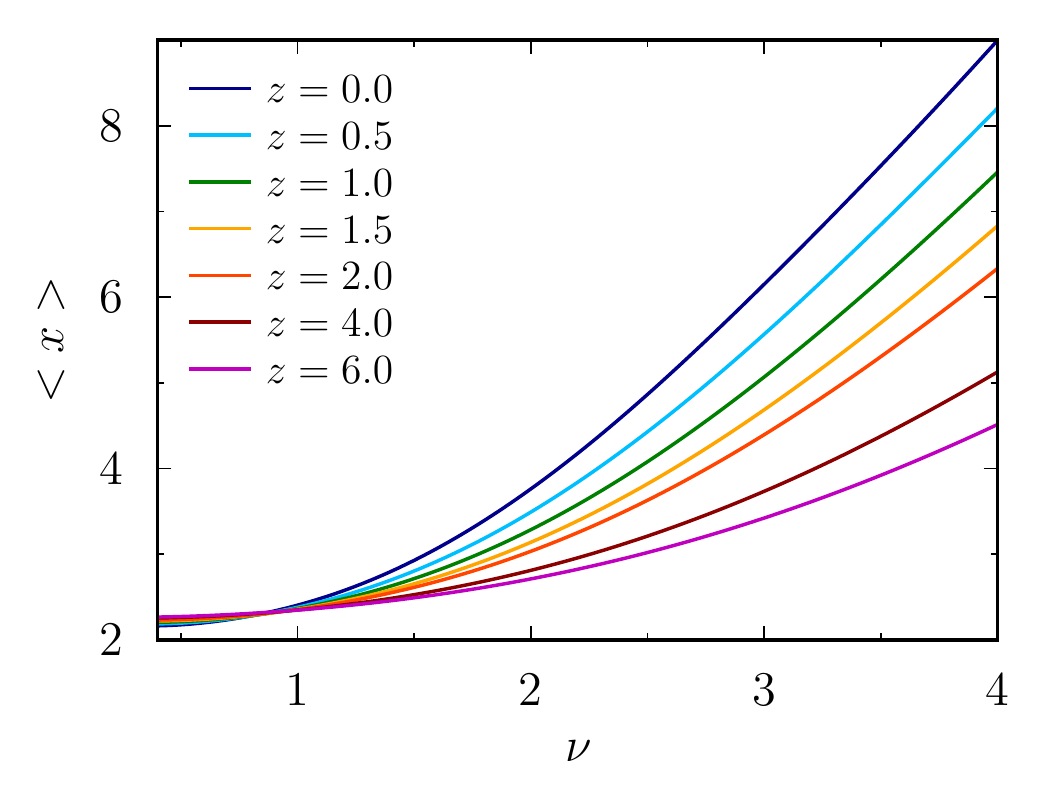}
\caption{Mean curvature, \xcurv, of peaks in a random Gaussian field for the power spectrum of the Bolshoi cosmology (Table \ref{table:cosmo}). While the curvature at fixed $\nu$ varies significantly between redshifts at high $\nu$, it cannot account for the non-universality of the \cnur at $\nu \approx 1$.}
\label{fig:curvature}
\end{figure}

In Section \ref{sec:results:model} we discussed how the slope of the matter power spectrum might influence concentration through two distinct physical effects: the abundance of sub--structure, and the shape of the peaks in the initial Gaussian random field. In this section, we describe our efforts to model the latter effect directly.

\citet{dalal_08} demonstrated that the shape of a peak in the initial Gaussian density field determines the mass accretion history of the halo it seeds; shallow parts of the peak will collapse quickly, while steeper parts will collapse slowly. Since the accretion history of a halo is intimately linked to its density profile and concentration \citep{ludlow_13_mah, ludlow_14_cm}, the shape of the initial peak has some bearing on the concentration of its offspring halo \citep{dalal_10}.

We can obtain a rough estimate of the ``steepness'' of a peak using the curvature parameter, defined as $x \equiv -\nabla^2 \delta / \sigma_2$ \citep{bardeen_86}. Here, $\delta$ is the overdensity field and $\sigma_2$ is the second moment of the variance (Equation (4.6c) in \citet{bardeen_86}, or Equation (\ref{eq:sigma}) with an extra $k^4$ factor in the integral). The higher moments such as $\sigma_2$ are ill--defined for the top--hat filter, and we thus switch to a Gaussian filter for this calculation. We compute the average curvature of peaks at fixed $\nu$, \xcurv, using the approximation given in Equation (6.14) of \citet{bardeen_86}. We have checked this approximation against the exact integral in Equation (A14) and found it to be accurate at the percent level.

Figure \ref{fig:curvature} shows \xcurv as a function of redshift and top-hat $\nu$. Due to its dependence on $k$, \xcurv differs with redshift at fixed $\nu$ and is thus a candidate for causing the non-universality of the \cnur. However, we note that at $\nu \approx 1$ the differences in \xcurv vanish, while it is precisely around this $\nu$ range that we observe  the largest deviations from universality of the \cnur (Figure \ref{fig:match_bolshoi}). Hence, the variations in \xcurv cannot by themselves account for the non-universality of the \cnur. 

This failure does not necessarily imply that peak curvature is not partly responsible for the non-universality in the \cnur. However, using curvature as one of the variables controlling concentrations would require at least one additional variable to explain the deviations from universality at low $\nu$. Furthermore, there are two possible reasons why \xcurv might not be the optimal parameter to consider. First, the mean curvature is computed for all peaks, but not all peaks form halos. In particular, small peaks are likely to be absorbed into larger halos (the so-called cloud-in-cloud problem; \citealt{bardeen_86}). Thus, the mean curvature in the low-$\nu$ regime does not correspond to the mean curvature of those peaks that end up forming halos. Second, \xcurv may not describe the slope of the outer profile accurately; for this reason, \citet{dalal_10} linked concentration to a particular measure of the outer slope of peaks as an alternative. Unfortunately, this measure suffers from the cloud-in-cloud problem as well.

Given these difficulties, the effect of peak shape would need to be directly measured in simulations, for example by tracing halos back to their initial Lagrangian volume and comparing their peak profile and concentration. However, the success of our model indicates that the dependence on the peak shape is already taken into account through the explicit dependence on the power spectrum slope. 

%--------------------------------------------------------------------------------------------
\section{Conversion to other mass definitions}
\label{sec:discussion:mdefs}
%--------------------------------------------------------------------------------------------

\begin{figure}
\centering
\includegraphics[trim = 1mm 18mm 2mm 81mm, clip, scale=0.75]{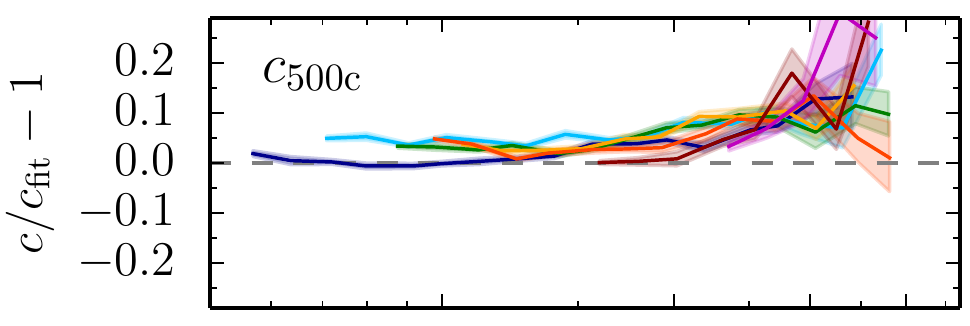}
\includegraphics[trim = 1mm 18mm 2mm 82mm, clip, scale=0.75]{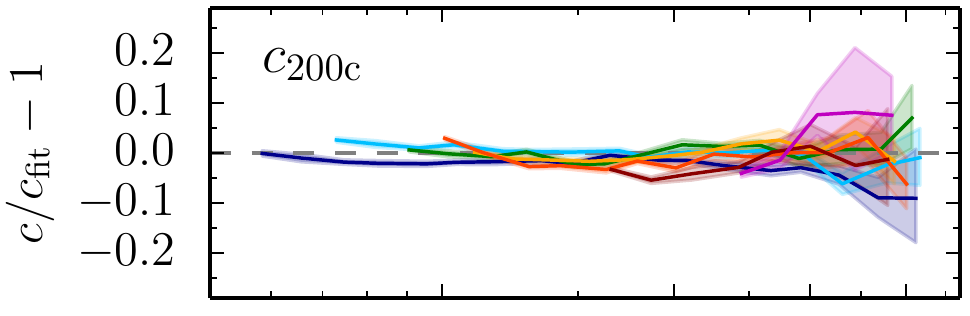}
\includegraphics[trim = 1mm 18mm 2mm 82mm, clip, scale=0.75]{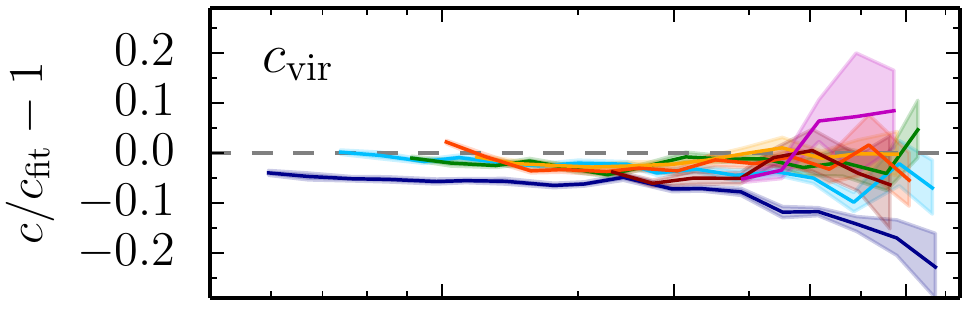}
\includegraphics[trim = 1mm 2mm 2mm 82mm, clip, scale=0.75]{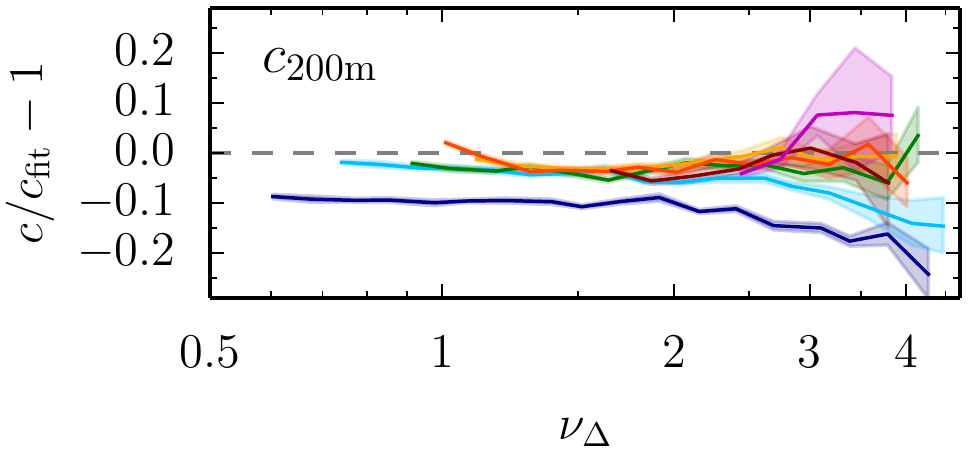}
\caption{Accuracy of the conversion of $\ctoc$ to other mass definitions, assuming NFW density profiles. The conversion degrades the accuracy of the model predictions by up to $\sim$ 15--20$\%$ at certain masses and redshifts. See text for details.}
\label{fig:mdefs_acc}
\end{figure}

The model proposed in Section \ref{sec:results:model} is based on the $\ctoc$ definition of concentration, and makes no direct prediction for the \cmr in other mass definitions. A conversion to other definitions can be performed a posteriori, but necessarily assumes a particular form of the density profile as a function of $\mtoc$ and $\ctoc$. If this functional form does not match the true density profile of halos, the \cmrs predicted for other mass definitions will deviate from those found in simulations. Figure \ref{fig:mdefs_acc} shows the accuracy of the \cnur for $\cfoc$, $\cvir$, and $\ctom$, as well as $\ctoc$ for comparison. The conversion from $\ctoc$ was performed assuming NFW density profiles. It is clear that the accuracy in the mass definitions other than $\ctoc$ is slightly degraded, though only in particular mass and redshift regimes. 

In the case of $\cfoc$, the difference with simulation results is caused by a systematic deviation of the NFW approximation from the real density profiles. Namely, at the highest peak heights, the profiles are steeper than predicted by the NFW profile, leading to a slightly underestimated $\rfoc$ (see, e.g., Figures 2 and 4 in \citealt{diemer_14_profiles}). The accuracy of the prediction can be improved to $\lsim 10\%$ when using the profile form suggested by \citet{diemer_14_profiles}. 

For $\cvir$ and $\ctom$, deviations appear only at low redshifts because $\rtom$, $\rvir$, and $\rtoc$ are almost the same radius at high redshift. At $\nu \lsim 2$ (roughly $10^{14} \msunh$ at $z = 0$), $\cvir$ is overestimated by $\sim 5\%$ and $\ctoc$ by $\sim 10\%$. At the highest peak heights, the differences increase to $\sim 15\%$ and $\sim 20\%$ for $\cvir$ and $\ctoc$, respectively. While using the \citet{diemer_14_profiles} profile improves the estimates by a few percent, the bulk of the effect is not caused by a deviation of the NFW profile from the true median profile. Instead, it appears that even if an accurate description of the mean or median profile is used in the conversion, the mean or median concentrations are biased high. Given the complex distributions of $c$ (Figure \ref{fig:distribution}), we have no reason to expect that a conversion based on the mean or median profiles should give perfect results. 

In conclusion, the conversion of our model to mass definitions other than $\ctoc$ introduces slight inaccuracies, but the concentrations still agree with simulation results to $\sim 10\%$. The only exception are definitions with large outer radii, such as $\ctom$, at high masses and low redshifts. In this particular regime, the error can increase to $\sim 20\%$. 

%%%%%%%%%%%%%%%%%%%%%%%%%%%%%%%%%%%%%%%%%%%%%%%%%%%%%%%%%%%%%%%%%%%%%%%%%%
% BIBLIOGRAPHY
%%%%%%%%%%%%%%%%%%%%%%%%%%%%%%%%%%%%%%%%%%%%%%%%%%%%%%%%%%%%%%%%%%%%%%%%%%

\bibliographystyle{apj}
\bibliography{sf}

\end{document}